\documentclass[manuscript]{aastex}
\usepackage{times}
\usepackage{helvet}
\usepackage{courier}
\usepackage{amssymb}
\usepackage{multirow}
\usepackage{amsopn}
\usepackage{epsfig}
\usepackage{graphicx,subfigure}
\usepackage{array}
\usepackage{url}
\usepackage{bm}
\usepackage{algorithm}
\usepackage{enumerate}
\usepackage{lscape}
\usepackage{times}
\usepackage{fancyhdr}
\usepackage{latexsym}
\usepackage{graphicx}
\usepackage{amsmath}
\usepackage{natbib}
\usepackage{upgreek}
\usepackage{pdflscape}
\usepackage{color}
\usepackage{ulem}
\usepackage{appendix}

\newcommand{\PreserveBackslash}[1]{\let\temp=\\#1\let\\=\temp}
\newcolumntype{C}[1]{>{\PreserveBackslash\centering}p{#1}}
\newcolumntype{R}[1]{>{\PreserveBackslash\raggedleft}p{#1}}
\newcolumntype{L}[1]{>{\PreserveBackslash\raggedright}p{#1}}

\newcommand{\ctwo}{$\rm C_2$}

\shorttitle{carbon stars}
\shortauthors{Y.B., Li, et al.}

\begin{document}

\title{Carbon stars identified from LAMOST DR4 using Machine Learning}

\author{
        Yin-Bi Li \altaffilmark{1},
        A-Li Luo$^{*}$ \altaffilmark{1},
        Chang-De Du \altaffilmark{1,2,3},
        Fang, Zuo \altaffilmark{1},
        Meng-Xin Wang \altaffilmark{1,2},
        Gang Zhao \altaffilmark{1},
        Bi-Wei Jiang \altaffilmark{4},
        Hua-Wei Zhang \altaffilmark{5},
        Chao Liu \altaffilmark{1},
        Li Qin \altaffilmark{1,2},
        Rui Wang \altaffilmark{1,2}
        Bing Du \altaffilmark{1,2},
        Yan-Xin Guo \altaffilmark{1,2},
        Bo Wang \altaffilmark{6},
        Zhan-Wen Han \altaffilmark{6},
        Mao-Sheng Xiang \altaffilmark{1,8},
        Yang Huang \altaffilmark{7},
        Bing-Qiu Chen \altaffilmark{7},
        Jian-Jun Chen \altaffilmark{1},
        Xiao Kong \altaffilmark{1,2},
        Wen Hou \altaffilmark{1},
        Yi-Han Song \altaffilmark{1},
        You-Fen Wang \altaffilmark{1}
        Ke-Fei Wu \altaffilmark{1,2},
        Jian-Nan Zhang \altaffilmark{1},
        Yong Zhang \altaffilmark{9},
        Yue-Fei Wang \altaffilmark{9},
        Zi-Huang Cao \altaffilmark{1},
        Yong-Hui Hou \altaffilmark{9}, and
        Yong-Heng Zhao \altaffilmark{1}
}
\altaffiltext{1}{Key Laboratory of Optical Astronomy, National Astronomical Observatories, Chinese Academy of Sciences, Beijing 100012, China;  lal@bao.ac.cn}
\altaffiltext{2}{University of Chinese Academy of Sciences, Beijing 100049, China.}
\altaffiltext{3}{Research Center for Brain-Inspired Intelligence, Institute of Automation, Chinese Academy of Sciences (CAS), Beijing 100190, China}
\altaffiltext{4}{Department of Astronomy, Beijing Normal University, Beijing 100875, China.}
\altaffiltext{5}{Department of Astronomy, School of Physics, Peking University, Beijing 100871, China.}
\altaffiltext{6}{Key Laboratory for the Structure and Evolution of Celestial Objects, Yunnan Observatories,Chinese Academy of Sciences, Kunming 650216, China.}
\altaffiltext{7}{South-Western Institute for Astronomy Research, Yunnan University, Kunming 650500, China}
\altaffiltext{8}{LAMOST Fellow}
\altaffiltext{9}{Nanjing Institute of Astronomical Optics $\&$ Technology, National Astronomical Observatories,Chinese Academy of Sciences, Nanjing 210042, China.}

\begin{abstract}

In this work, we present a catalog of 2651 carbon stars from the fourth Data Release (DR4) of the Large Sky Area Multi-Object Fiber Spectroscopy Telescope (LAMOST). Using an efficient machine-learning algorithm, we find out these stars from more than seven million spectra. As a by-product, 17 carbon-enhanced metal-poor (CEMP) turnoff star candidates are also reported in this paper, and they are preliminarily identified by their atmospheric parameters. Except for 176 stars that could not be given spectral types, we classify the other 2475 carbon stars into five subtypes including 864 C-H, 226 C-R, 400 C-J, 266 C-N, and 719 barium stars based on a series of spectral features. Furthermore, we divide the C-J stars into three subtypes of C-J(H), C-J(R), C-J(N), and about 90$\%$ of them are cool N-type stars as expected from previous literature. Beside spectroscopic classification, we also match these carbon stars to multiple broadband photometries. Using ultraviolet photometry data, we find that 25 carbon stars have FUV detections and they are likely to be in binary systems with compact white dwarf companions.

\end{abstract}


\section{Introduction}

Carbon stars were first recognized by \citet{secchi1869}, who loosely defined these stars as those with optical spectra dominated by carbon molecular bands, such as CN, CH, or the Swan bands of C$_{2}$. They were initially considered to be asymptotic giant branch (AGB) stars because the carbon element in their atmosphere can only be dredged to the surface during the third dredge-up process. Researchers did not realize that an extrinsic process can also produce carbon stars until the first main-sequence carbon star (dC) G77-61 was discovered by \citet{dahn1977}. \citet{dearborn1986} pointed out that G77-61 is a single-line binary and has a cool optical invisible white dwarf companion. Such binary theories hypothesize that carbon-rich material may be lost through the stellar wind from a thermally pulsing AGB star, and accreted by its companion main-sequence star \citep{han1995}.

Numerous abundance studies of metal-poor stars \citep{beers1985, beers1992, wisotzki1996, Christlieb2001, Christlieb2003} have discovered the carbon-enhanced metal-poor (CEMP) stars, which were originally defined as stars with metallicity $[\rm Fe/H] <= -1.0$ and $[\rm C/Fe] >= 1.0$ \citep{beers2005}. Among the CEMP stars, main-sequence turnoff stars are expected to be of particular importance because they might preserve on their surfaces pure material accreted from their AGB companions, which can be used to investigate the production efficiency of carbon and neutron-capture elements of AGB stars.

\subsection{Types of Carbon Stars}

\citet{keenan1993} proposed the Morgan-Keenan (MK) classification system to divide carbon stars into seven types based on significant spectral features, and it was widely applied to the subsequent classification for medium- or low-resolution spectra of carbon stars \citep{wallerstein1998, barnbaum1996, goswami2005, goswami2010, lloyd2010}. \citet{barnbaum1996} revised the MK classification criteria, which are listed in Table~\ref{tab: classification_criteria}. We adopt this revised MK system to classify carbon stars into five types, i.e., C-H, C-R, C-J, C-N, and Barium stars, in this paper.

C-H stars are known to be binaries with a compact white dwarf companion \citep{mcclure1983, mcclure1984, mcclure1990}, and they can be recognized from the dominance of the secondary P-branch head near 4342 $\rm\AA$ as listed in Table~\ref{tab: classification_criteria}. C-R stars were once considered to be binaries, but are now taken to be single stars coming from binary mergers \citep{izzard2007, dominguez2010}. They can be identified by the almost equal depths of the Ca I line at 4226 $\rm\AA$ and the CN band at 4215 $\rm\AA$ \citep{goswami2005, goswami2010}. C-N stars are AGB stars,  and can be distinguished by the predominant feature of a blue depression often nearly obliterating the spectrum below 4400 $\rm\AA$ \citep{barnbaum1996, goswami2005, goswami2010}. Like the C-R stars, the exact evolutionary stage and nature of C-J stars are still unclear \citep{abia2003}. Their spectra possess unusually strong isotopic carbon bands, and are clearly recognized by having a j index larger than 4 \citep{demello2009}. Barium stars are red giants showing strong spectral lines of s-process elements, especially Ba~II at 4554 $\rm\AA$ and 6496 $\rm\AA$, and Sr~II at 4077 $\rm\AA$, which make them easily recognizable \citep{de Castro2016}.

\subsection{Carbon Star catalogs in the Literature}

As we know, carbon-enhanced stars play important roles in understanding the evolution of the Galaxy, but the number of these stars with detailed chemical abundances studies is small. To date, a series of works have been conducted to search for Galactic carbon stars, and we summarize them in Table~\ref{table:previous_survey_1}.

In 2001, \citet{alksnis2001} published a revised catalog containing 6891 carbon stars, which is based on a collection of journal articles \citet{stephenson1973, stephenson1989}. \citet{Christlieb2001} presented a sample of 403 faint high-latitude carbon (FHLC) stars from the digitized objective prism plates of the Hamburg/ESO Survey (HES). From the years 2002 to 2013, carbon stars were systematically searched from the Sloan Digital Sky Survey (SDSS). \citet{margon2002}, \citet{downes2004}, and \citet{green2013} respectively found 39, 251, and 1220 FHLC stars. In addition, \citet{sijianmin2014} was the first to apply a machine learning algorithm to SDSS DR8 spectra and discovered 260 new carbon stars. The Large Sky Area Multi-Object Fiber Spectroscopy Telescope (LAMOST; Wang et al. 1996; Su \& Cui 2004; Zhao et al. 2012; Luo et al. 2012; Cui et al. 2012) began to release data from the year of 2011; new carbon stars were reported from massive number of LAMOST spectra. \citet{sijianmin2015} applied a manifold ranking algorithm to Pilot survey data and obtained 183 new carbon stars. \citet{jiwei2016} identified 894 carbon stars from LAMOST DR2 with multiple spectral line indices.

\subsection{The Rank-based PU Learning Algorithm}

LAMOST has obtained 7,725,624 spectra in the first four years of regular survey, and searching for carbon stars from such a massive dataset is our main aim. Since carbon stars are extremely rare, it is impossible to manually seek a small number of carbon stars from the massive data. We turned to machine-leaning (ML) methods, and finally adopted the Bagging TopPush algorithm \citep{duchangde2016} to retrieve carbon stars from LAMOST DR4. It is a supervised rank-based PU learning algorithm and needs positive and negative samples to train the rank model. In our work, carbon stars reported by previous works can be used as a positive sample set (P), and the massive LAMOST data can be treated as unlabeled samples (U). The Bagging TopPush algorithm randomly selects negative samples from the set U and develops models using the positive and negative samples at first, ranking the positive samples ahead of the selected negative samples. Then, the developed model calculates the scores for all unlabeled samples in set U and rank them in descending order by their scores. In this ranked unlabeled sample list, the algorithm should rank carbon stars ahead of other objects.

This paper is organized as follows. In Section 2, we introduce the steps to select positive samples and the method to cluster them. In Section 3, we firstly analyze the effect of spectral preprocessing methods on algorithm performance and then determine the spectral preprocess method in sub-section 3.1. Then, we search for carbon stars from LAMOST DR4, and roughly estimate the completeness and contamination in sub-section 3.2. Next, we analyze the classification results of our carbon stars given by the LAMOST 1D pipeline in sub-section 3.3, and compare our stars with two previous catalogs in sub-section 3.4. In the end, we also find 17 carbon-enhanced metal-poor turnoff (CEMPTO) stars, and preliminarily verify their nature as such by measuring their atmospheric parameters from low-resolution spectra in sub-section 3.5. In Section 4, we classify carbon stars into five types, and analyze their spatial and magnitude distributions. In Section 5, we investigate carbon stars using photometric data, for example, the distributions of ultraviolet, optical and infrared magnitudes. Finally, a brief summary is given in Section 6.

\section{Positive Samples}
\citet{duchangde2016} investigated in detail the performance of six widely used machine-learning algorithms, and pointed out the Bagging TopPush algorithm has the best performance, lowest computation complexity and is least CPU time consuming. Three parameters of the Bagging TopPush algorithm need to be determined before using it, and \citet{duchangde2016} had studied their value ranges through comparing performances of the algorithm under different parameter values. Here we inherit Du¡®s parameter and only need to construct positive sample set P.

\subsection{Positive Sample Selection}

In this sub-section, we select positive samples from previous literature that provide catalogs of carbon stars from SDSS and LAMOST. \citet{margon2002} and \citet{downes2004} respectively reported 39 and 251 carbon FHLCs from the commissioning data and the First Data Release (DR1) of SDSS, and \citet{green2013} found 1220 FHLCs from the Seventh Data Release (DR7) of SDSS, which is about five times greater than previously found. Furhtermore, \citet{sijianmin2015} found 183 carbon stars from the Large Sky Area Multi-Object Fiber Spectroscopic Telescope (LAMOST) pilot survey---158 of them were reported for the first time---and is the first research work of carbon stars using LAMOST spectra.

We obtain the equatorial coordinates of the SDSS carbon stars reported in the above three publications from the SIMBAD Astronomical Database\footnote{http://simbad.u-strasbg.fr/simbad/} and download their spectra from the Web site of SDSS DR12\footnote{http://dr12.sdss3.org/bulkSpectra}. For stars with multiple observations, after submitting their coordinates to DR12, we only download the spectrum with the highest signal-to-noise ratio (S/N) for each star. For 183 LAMOST carbon stars, we download their spectra from the Web site of LAMOST DR1\footnote{http://dr1.lamost.org/}. In total, we obtain 1682 carbon star spectra from SDSS and the LAMOST survey, and we further select positive samples through the following four steps, with the numbers in the parentheses indicating the amount of stars left after each step.

\begin{enumerate}
  \item For stars observed several times, we only retain the spectrum with the highest S/N, and remove other repeated spectra. Through this step, we remove 234 SDSS spectra (1448).
  \item In our experience, the S/Ns of the blue and red ends in an SDSS or LAMOST spectrum are generally low, thus we restrict our spectra to begin at 3920 \AA to have 3590 wavelength points. There exists a few spectra that have different wavelength ranges from our limitation, thus we remove 13 spectra, which start from wavelengths larger than 3920 \AA (1435).
  \item For SDSS spectra, there may be no flux information within a fraction of a wavelength. We individually check all SDSS carbon star spectra, and remove such spectra. For LAMOST spectra, they may have zero fluxes which were artificially set to represent negative fluxes, may have low quality in the red and blue overlaps in the range of [5700 \AA, 6000 \AA], and may exhibit the water line around 7600 \AA. We also check all LAMOST spectra and rule them out. In this step, 52 SDSS spectra and 49 LAMOST spectra are removed (1334).
  \item For a fraction of carbon stars, their entire spectra are contaminated by considerable noise, and they have extremely low S/Ns. When we manually inspect them, we find that it is difficult for us to identify them as carbon stars, thus we exclude these spectra. Through this step, seven LAMOST spectra and 277 SDSS spectra are removed (1050).
\end{enumerate}

After these four steps, we totally obtain a total of 127 LAMOST spectra and 923 SDSS spectra, and treat them as positive samples to construct the positive sample set P. In appendix A, we respectively show the S/N distributions of the 1334 spectra left after the above step three and 1050 positive sample spectra. In addition, we also preliminarily analyze the effect of these extreme low S/N spectra on algorithm performance.

\subsection{Positive Sample Clustering}

\begin{figure}
\centering
\epsscale{1.0}
\plotone{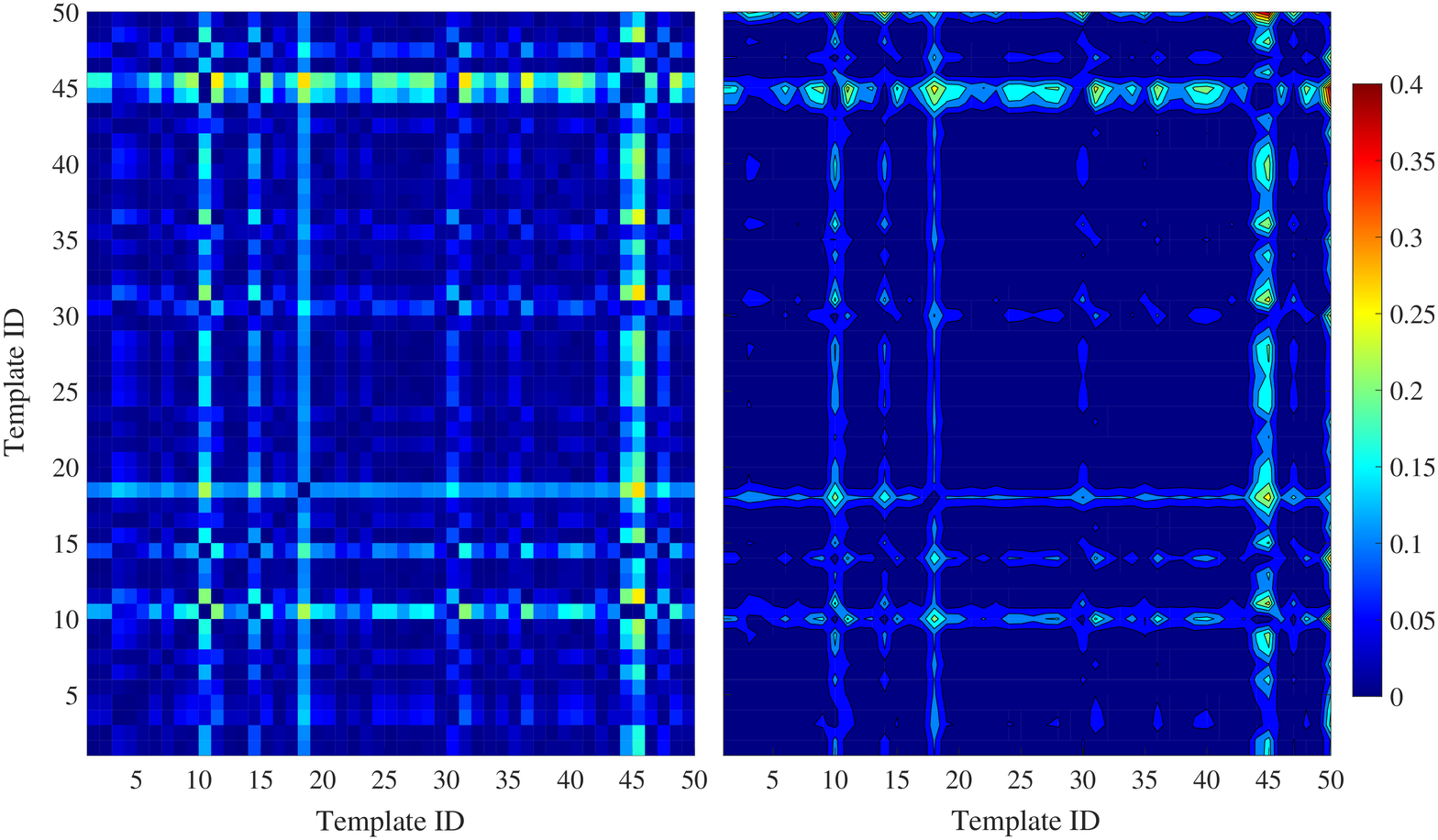}
\caption{(Left) Number density distribution of the cosine distances of each of the two templates. (Right) Contour plot of these distances.
\label{number_density_contour}}
\end{figure}

The number of cool C-N and C-J stars in our 1050 positive samples is much smaller than that of other star types. In this case, a fraction of cool carbon stars in the massive LAMOST data are difficult to rank into the top $K$ results if we use the Bagging TopPush method. Thus, we need to divide the 1050 positive samples into different clusters and retrieve the carbon stars with these clusters. Furthermore, carbon stars can be classified into five types using the revised MK classification system \citep{barnbaum1996} as mentioned in the introduction. But, the MK types of the 1050 samples have not been provided in the literature, and it is necessary for us to cluster positive samples first. We adopt the unsupervised $K$-mean cluster algorithm \citep{seber1984, spath1985} to classify the 1050 positive samples, but we do not know how many groups can be classified in advance. Thus, we adopt the commonly used approach to classify positive samples into tens or even hundreds of groups at first, and then combine groups with high similarities. Initially, we cluster the 1050 samples are clustered into 50 groups. In this step, we randomly select 50 positive samples as the initial cluster centers, and several test experiments show that the classification results are stable and do not depend on the selection of the initial cluster centers. Then, we estimate the cosine distance of the cluster center spectra of each of the two groups; the number density distribution and contour plot of these cosine distances are depicted in Figure~\ref{number_density_contour}.

\begin{figure}
\centering
\epsscale{1.0}
\plotone{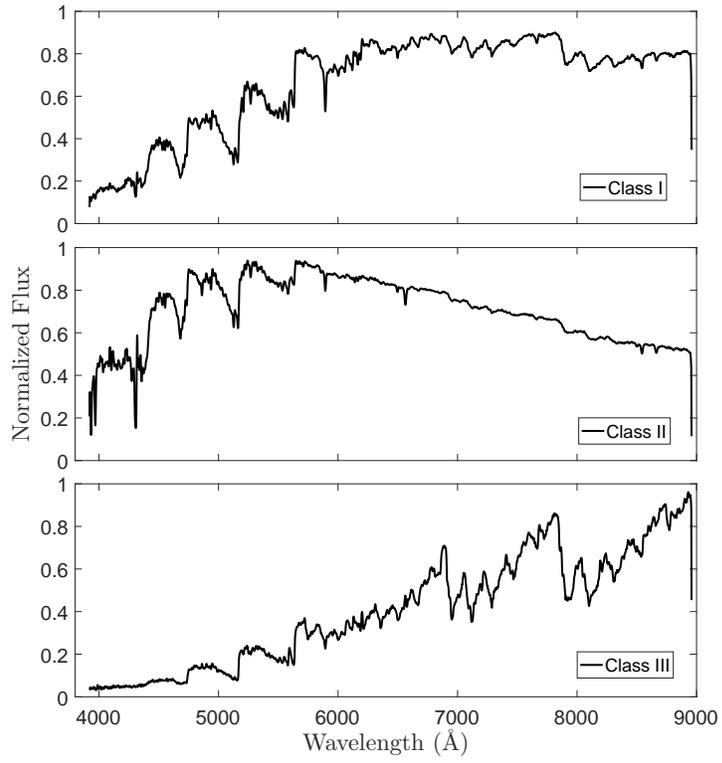}
\caption{Cluster center spectra of positive samples for the three types. The horizontal axis is the wavelength, and the vertical axis is the normalized flux.
\label{three_type_templates}}
\end{figure}

From this figure, we manually selected nine outlier groups, shown by the lighter colors, and then we divide our 50 groups of positive samples into two groups. Forty-one groups form the first cluster (C$\_$I), and the nine outlier groups form the second cluster (C$\_$II). We calculate the inner average cosine distances of C$\_$I and C$\_$II (D$\_$I and D$\_$II), and also their outer distance (D$\_$I$\_$II). D$\_$I and D$\_$II are respectively 0.0082 and 0.0324, and D$\_$I$\_$II is 0.0467. We found that D$\_$I$\_$II is not much larger than D$\_$II, and the inner distance D$\_$II is too large. Thus, we try to divide the C$\_$II into two clusters (C$\_$II and C$\_$III), still using the $k$-mean method, and estimate the inner and outer distances. The inner average distances of the three clusters are 0.0088(D$\_$I), 0.0081(D$\_$II), and 0.0165(D$\_$III) respectively, and their  outer distances are 0.0613 (D$\_$I$\_$II), 0.0911 (D$\_$I$\_$III), and 0.2585 (D$\_$II$\_$III). We can see that their inner distances are much less than their outer distances if we divide the positive samples into three clusters, which ensures that their inner distances are as small as possible and their outer distances are as large as possible. Thus, we determine that our 1050 positive samples can be divided into the three clusters. There are a total of 743 positive samples classified as C$\_$I, 278 positive samples belonging to C$\_$II, and 29 positive samples classified as C$\_$III. Figure~\ref{three_type_templates} shows the spectra of the cluster centers with normalized fluxes.

Such a classification result is different from the spectral subtypes of carbon stars (five types), but it is understandable. The C-H, C-R, and barium stars, they lie in a smaller color range, which indicates that they have similar effective temperatures and spectral types. Thus, their main spectral features are similar, for example, the continuum and carbon molecular band features. As mentioned in the abstract, the method to distinguish them mainly depends on the spectral line features. For example, Ba stars can be distinguished from C-R stars based on the strong spectral lines of Ba II at 4554 $\rm\AA$ and Sr II at 4077 $\rm\AA$. Such subtle features would be ignored by the similarity measures, which only focus on main features. For cool C-J and C-N stars, the same problem exists. The difference between them only lies in the strong isotropic carbon line and band, thus they will be clustered into the same sub-type with a similar continuum and strong CN bands. Thus, the similarity measure will put different spectra subtypes into the same groups depending on the main spectral features, and will ignore the subtle features that may be useful for the classification of C-H, C-R, C-J, C-N, and Ba stars.

In the next section, we will retrieve carbon stars with the three types of positive samples, and combine their results.

\section{LAMOST Carbon Stars}

\subsection{Determination of Spectral Preprocessing Method}

Before using the Bagging TopPush algorithm to retrieve carbon stars, preprocessing methods for spectra are investigated to determine both the effect of spectral preprocessing methods on algorithm performance, and determine the method used in this paper. For positive samples of each group, we adopt eight methods, which are detailed in Appendix B, to preprocess the spectra. For each preprocessing method, we execute a retrieving experiment using the Bagging TopPush method \citep{duchangde2016}, and compare the experiment results. There are two approaches to displaying the experiment results: one is the relationship of recall and $K$, and the other is the relation between precision and $K$. Recall ($n/N$) is the fraction of retrieved carbon stars ($n$) within all carbon stars of the unlabeled sample set U ($N$), and the precision ($n/K$) represents the proportion of retrieved carbon stars ($n$) within the top $K$ ranking samples ($K$) in descending order.

The experiment results are  shown in Figure ~\ref{preprocess_I}--\ref{preprocess_III}, respectively. The left panels in the three figures are the relationships of recall and $K$, and the right panels are the variation of precision with $K$. In these figures, the eight color polylines represent eight preprocessing methods: ``nmap,'' ``nunit,'' ``nmap + pca,'' ``nunit + pca,'' ``nmap + subcon,'' ``nunit + subcon,'' ``nmap + pca + subcon,'' and ``nunit + pca + subcon.'' Among these, the ``nmap'' and ``nunit'' are two spectral normalization methods, while ``subcon''  and ``pca'' stand for the continuum subtraction and principal component analysis (PCA) method, respectively.

The figures tell us that the preprocessing approach named as ``nmap + pca'' shown in purple solid polylines, has both the highest recall and precision over the others. Thus, in this paper, we conspicuously employ the combination of ``nmap'' plus ``pca'' as our pre-process method for all spectra. The symbol ``nmap + pca'' means that we use the ``nmap'' method to normalize spectra and apply the PCA method to reduce the data dimension.

Because we cannot know definitely the number of carbon stars in the massive LAMOST DR4 dataset, thus it is impossible to accurately estimate the completeness and contamination rates of our method. However, we can roughly evaluate them using above preprocessing experiments. For the three groups of positive samples, the recall of the ``nmap + pca'' method at $K = 5000$ ($r1$, $r2$, and $r3$) is respectively 99.54$\%$, 99.78$\%$, and 100$\%$, and the precision at $K = 5000$ ($p1$, $p2$, and $p3$) is respectively 7.39$\%$, 2.77$\%$, and 0.28$\%$. As a result, the recall can be roughly treated as the completeness rates for the three groups of positive samples, and $1 - p1$, $1 - p2$ and $1 - p3$ are their contamination rates, respectively. We can see that the completeness at $K = 5000$ is extremely high; meanwhile the contamination at the $K = 5000$ is also very high. However, for the retrieval of carbon stars from the massive spectral data set, we are more concerned about the completeness, but hardly care about the contamination. This is because the subsequent works can help us to decrease the contamination rate, for example, through repeat retrieval and manual inspection and identification.

\subsection{Carbon Stars from LAMOST DR4}

In this subsection, we adopt the Bagging TopPush approach \citep{duchangde2016} to retrieve carbon stars from LAMOST DR4. During the first four years of the LAMOST regular survey, 20$\%$ of the nights in each month were used for instrument tests, and spectra obtained in such test nights were not published. In order to find more carbon stars, we use both released and test night spectra. Taking into account the computer memory capacity, we divide all LAMOST spectra into 24 groups, and there are 500,000 spectra included in each group. We use the positive sample set P1 to retrieve carbon stars from each 500,000 LAMOST spectra, then return the top 5000 spectra of ranked results. For 24 group LAMOST spectra, we obtain 120,000 preliminary candidates. We then continue to use the P1 set to retrieve carbon star from the 120,000 samples, and output the top 20,000 spectra for automated identification. For the other two positive sample sets P2 and P3, which have 278 and 29 spectra respectively, we also perform a similar procedure. Finally, we combine results of the three group positive samples together, and there are a total of 60,000 spectra which need visual inspection.

After a manual check, we find over 3000 carbon star spectra. Since a fraction of objects were observed several times, we only retain the spectrum with the highest S/N, and finally identify 2651 carbon stars from LAMOST DR4. Comparing with previous catalogs, we find that 1415 of them have not been reported in previous works. Moreover, among these carbon stars, 296 of them were observed in test nights, and thus their spectra were not published on the Web site of LAMOST DR4\footnote{http://dr4.lamost.org/}.

\subsection{Comparison with the LAMOST 1D Pipeline}

For our carbon stars, it is essential to obtain the number of stars classified correctly by the spectral analysis software of LAMOST (LAMOST 1D pipeline), which can be used to determine whether the machine-learning algorithm adopted in this work shows good performances. We obtained the spectral types given by the LAMOST 1D pipeline for our carbon stars and list them in Table~\ref{table:1D classification results}. From this table, we can clearly see that 1671 (about 63$\%$) of our carbon stars were classified correctly, but others (37$\%$) were incorrectly classified by the LAMOST 1D pipeline. Among these incorrectly classified carbon stars, 74$\%$ (726) of them were classified as G-type stars (mainly G5 type stars), and others were respectively categorized as DQ white dwarf (two), an A9-type star (one), F-type stars (four), K-type stars (97), M-type stars (one), and unknown objects (149).

\begin{figure}
\centering
\epsscale{1.0}
\plotone{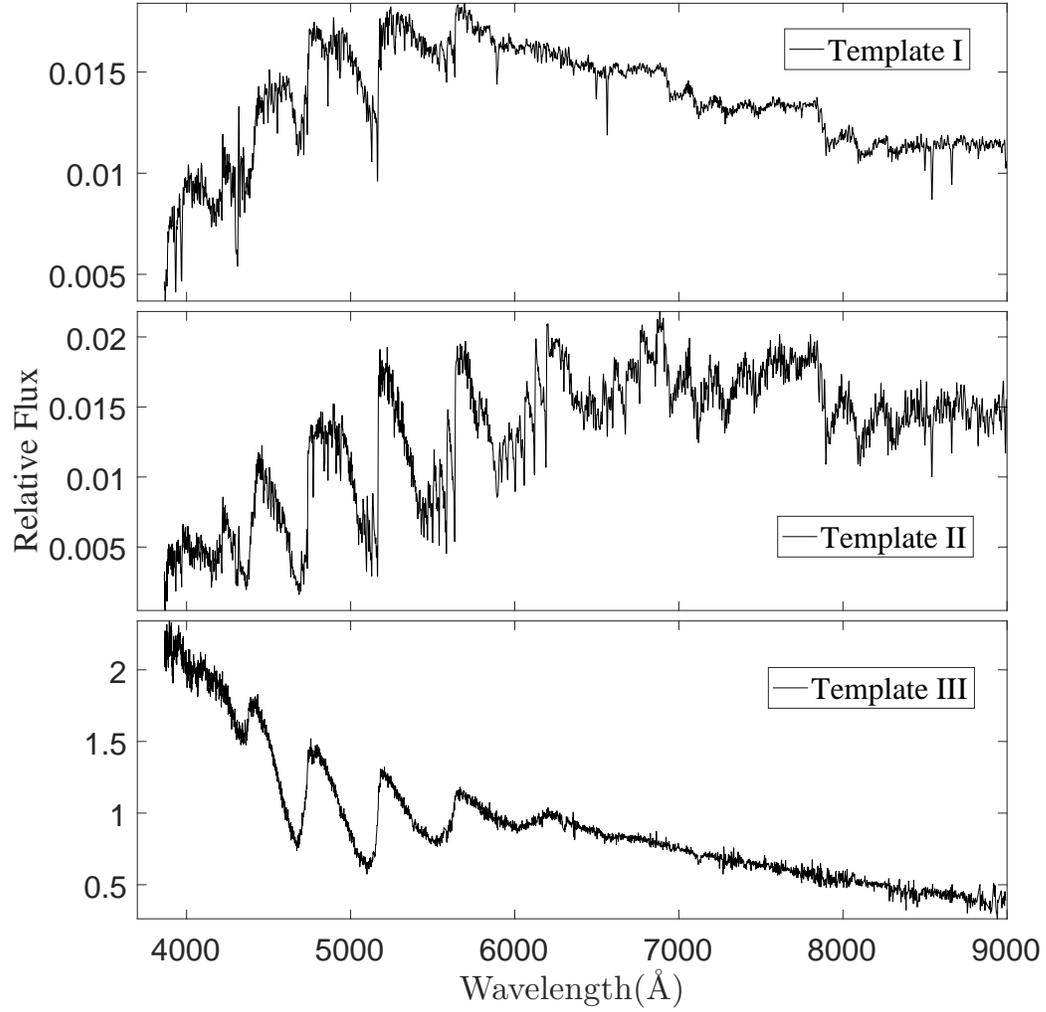}
\caption{The three template spectra for carbon stars used by the LAMOST 1D pipeline. \label{1d_templates}}
\end{figure}

Compared with that for F-, G- or K-type stars, the classification accuracy of the LAMOST 1D pipeline for carbon stars is relatively low, which is probably arose from carbon stars having fewer template spectra. Figure~\ref{1d_templates} shows the carbon star templates used by the LAMOST 1D pipeline now. We can see that the LAMOST 1D pipeline has three templates for carbon stars, two of which are common carbon stars and the last one is a DQ white dwarf. Obviously, the two carbon star templates are relatively hot stars, and no cool C-N type carbon star is included in the templates. In addition, a few types of carbon stars which have special spectral features, such as CEMP stars and carbon subdwarf stars, are also not included in the templates. Thus, constructing new carbon star template spectra will be an important work in the future for the LAMOST 1D pipeline, which will greatly increase classification accuracy and reliability.

\subsection{Comparing with Previous Works}

Before this work, \citet{sijianmin2015} found 183 carbon stars from the LAMOST pilot survey using an efficient machine-learning method, and \citet{jiwei2016} reported 894 carbon stars from LAMOST DR2 using a series of molecular band index criteria. Thus, it is important to compare our carbon stars with these two publications.

We compare our stars with the two previous catalogs of carbon stars \citep{sijianmin2015, jiwei2016}, and the results listed in Table~\ref{table: si_ji}. The first, second, and third columns respectively give the references, the total number of stars and the data used by the two catalogs. The fourth column, $N\_{1}$, gives the number of carbon stars in two previous catalogs that was eliminated in our catalog during manual identification because of low spectral quality. The fifth column, $N\_{2}$, shows number of carbon stars included in our catalog but excluded from previous catalogs. We can see that there are a total of 33 and 575 stars excluded from the catalog of \citet{sijianmin2015} and \citet{jiwei2016} respectively. The two publications used two older versions of the data set, which did not include data obtained in test nights. In contrast, we used the latest version of LAMOST spectra and data observed during the test nights. Furthermore, we also use a more efficient method to find carbon stars. These are the reasons why there are relatively large numbers of carbon stars omitted by the two catalogs. The sixth column, $N\_{\rm test}$, shows the number of carbon stars, which were obtained during the test nights but excluded by previous literature, and there are respectively four and 122 carbon stars omitted because test night data were not used. The seventh column, $N\_{\rm version}$, provides the number of carbon stars ignored by the two catalogs since the old version data were used. The last column, $N\_{\rm algorithm}$, lists the number of carbon stars removed by the selection methods in the two publications. There are 17 stars eliminated from the catalog of \citet{sijianmin2015} by the machine-learning method (EMR) they used. \citet{duchangde2016} compared the EMR algorithm with our Bagging TopPush method, and found that the performance of Bagging TopPush is better than that of the EMR algorithm. Further, there were 403 carbon stars excluded from the work of  \citet{jiwei2016} by their target selection method. We use each step of their selection criteria to check how these stars were eliminated, and list the results in Table~\ref{tab:steps}.

From this table, we can see that the first step is an S/N criterion, which omits 32 carbon stars from our catalog. We carefully check the SNRs of these 32 spectra, and find their spectral qualities are good enough to be recognized visually. The S/Ns provided by the LAMOST pipeline are calculated based on continua, and it is just a rough indicator instead of an exact quality measurement of a spectrum. Besides, the $i$-band S/Ns used in this step are not the best spectral quality estimates for the late-type stars. Then, steps 2, 3, 4, and 5 adopted molecular band index selection criteria, and 25, 138, 113, and 54 carbon stars were ignored in each of the four steps, respectively, resulting in a total of 330 carbon stars being missed by these molecular band index methods. Finally, 41 carbon stars were removed by a color selection criterion in step 6, which could eliminate carbon stars with hotter temperatures. We check the spectra of these 41 stars, and all of them are hot carbon stars as expected, which means that relatively strict color criteria could exclude a small fraction of carbon stars.

\subsection{Carbon-enhanced Main-sequence TurnOff Star Candidates}

\begin{figure}
\centering
\epsscale{1.0}
\plotone{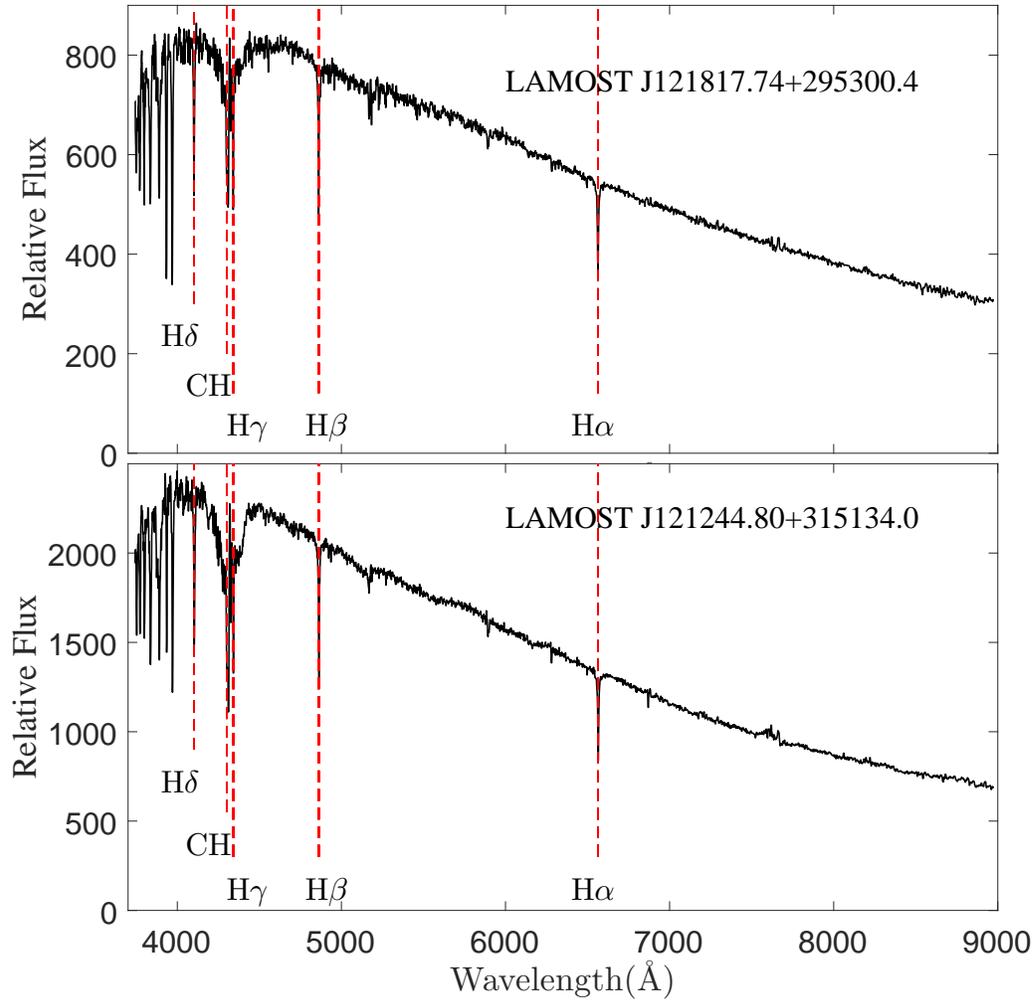}
\caption{Two example spectra of the CEMP turnoff stars, which have strong Balmer absorption lines and CH molecular bands. \label{spectra_cemptf}}
\end{figure}

In addition to the above 2651 carbon stars, we also find 17 stars that have strong Balmer absorption lines and C-H molecular bands in their spectr, as shown in Figure~\ref{spectra_cemptf}; these features show that they are hot carbon-enhanced stars and have higher effective temperatures than other types of carbon stars. We check their atmospheric parameters given by the LAMOST stellar parameter pipeline (LASP) in our database, but the LASP does not provide any. Thus, we estimate their effective temperatures, surface gravities, and metallicities with the $\chi^{2}$ minimization technique described in the sub-section 4.4 of \citet{lee2008}, and list them in Table~\ref{table:basic_parameters_cempto}, which also includes other basic information, i.e., equatorial coordinates, $r$-band signal-to-noise ratio ``S/N$\_{\rm r}$'' and spectral type ``SpType$\_{\rm PL}$'' given by the LAMOST 1D pipeline. From this table, we can see that their effective temperatures are all greater than 5800, and their surface gravities are all greater than 3.6 except one, which indicates that they are likely located at the main-sequence turnoff region on the Hertzsprung-Russell diagram. In addition, their metallicities, parameterized by [Fe/H], are all below -1.6 and even as low as -2.4 except for one star, indicating that they are metal-poor stars. Thus, we infer that these 17 stars are likely to be CEMPTO stars as mentioned in \citet{aoki2008}, and need future high-resolution follow-up observation for identification.

We cross-match these 17 stars with the Galaxy Evolution Explorer (GALEX; \citep{martin2005}), Panoramic Survey Telescope and Rapid Response System (Pan-STARRS; \citep{kaiser2002, kaiser2010, chambers2016}), The Two Micron All Sky Survey (2MASS; \citep{skrutskie2006}), and Wide-field Infrared Survey Explorer (WISE; \citep{wright2010}) which will be described in section 5, and list their ultraviolet, optical, and infrared magnitudes in Table~\ref{table:magnitudes_cempto}. From this table, we note that none of them have an FUV detection, suggesting that it is highly likely that none of them are in a binary system with a compact white dwarf companion. In addition, their radial velocities ``RV'' given by the LAMOST 1D pipeline and proper motions in the right ascension and declination directions from the PPMXL \citep{roeser2010} and UCAC4 \citep{zacharias2013} catalogs are listed in Table~\ref{table:motion_cempto}, where ``$\mu_{\alpha}\cos(\delta)_{\_ \rm P}$'' and ``$\mu_{\delta}$$_{\_ \rm P}$'' are the two proper motions from the PPMXL catalog, and ``$\mu_{\alpha}\cos(\delta)_{\_ \rm U}$'' and ``$\mu_{\delta}$$_{\_ \rm U}$'' are the proper motions from the UCAC4 catalog. From this table, we can see that all of them have PPMXL proper motions but 13 of them have UCAC4 proper motions. The two proper motions are roughly consistent, and most of them have large proper motions, consistent with their possible nature as main-sequence turnoff stars.

\section{Spectral Types and Spatial Distribution}

\subsection{Classification}

In this paper, we adopt the \citet{barnbaum1996} classification system to categorize our 2651 carbon stars; their prominent spectral features are  summarized in table \ref{tab: classification_criteria} and are detailed in the introduction. Based on these criteria, we classify the 2651 carbon stars into five types.

\begin{figure}
\centering
\epsscale{1.0}
\plotone{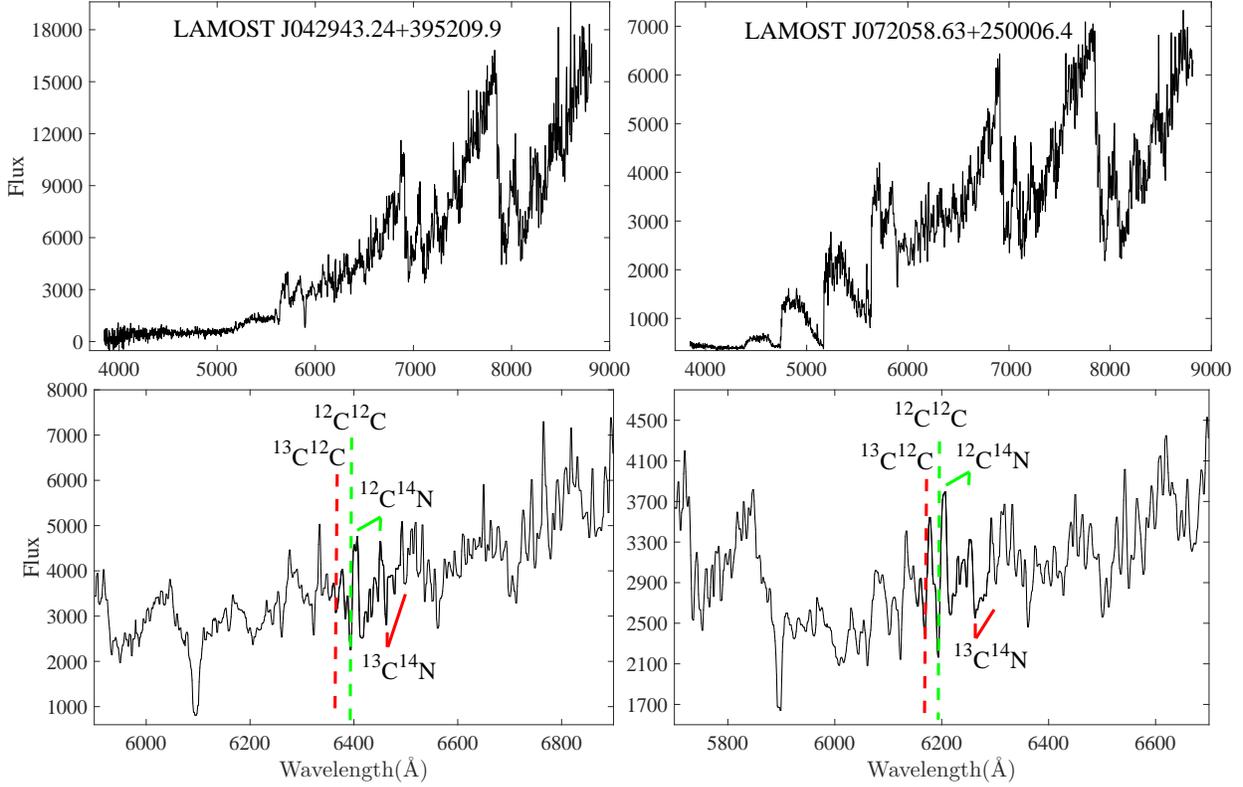}
\caption{The upper panels are two spectra of C-J(N)-type stars, and the bottom panels are their local spectra from 5700 to 6700 $\rm\AA$, used to calculate their pseudo-continua and further $j$ indexes.
The spectral lines of $^{13}$C$^{12}$C at 6168 $\rm\AA$ and $^{12}$C$^{12}$C at 6192 $\rm\AA$ are, respectively, marked by red and green dashed lines, and the normal $^{12}$C$^{14}$N bands at 6206 $\rm\AA$ and
isotopic $^{13}$C$^{14}$N bands at 6260 $\rm\AA$ are also displayed. \label{cj_n}}
\end{figure}

\begin{figure}
\centering
\epsscale{1.0}
\plotone{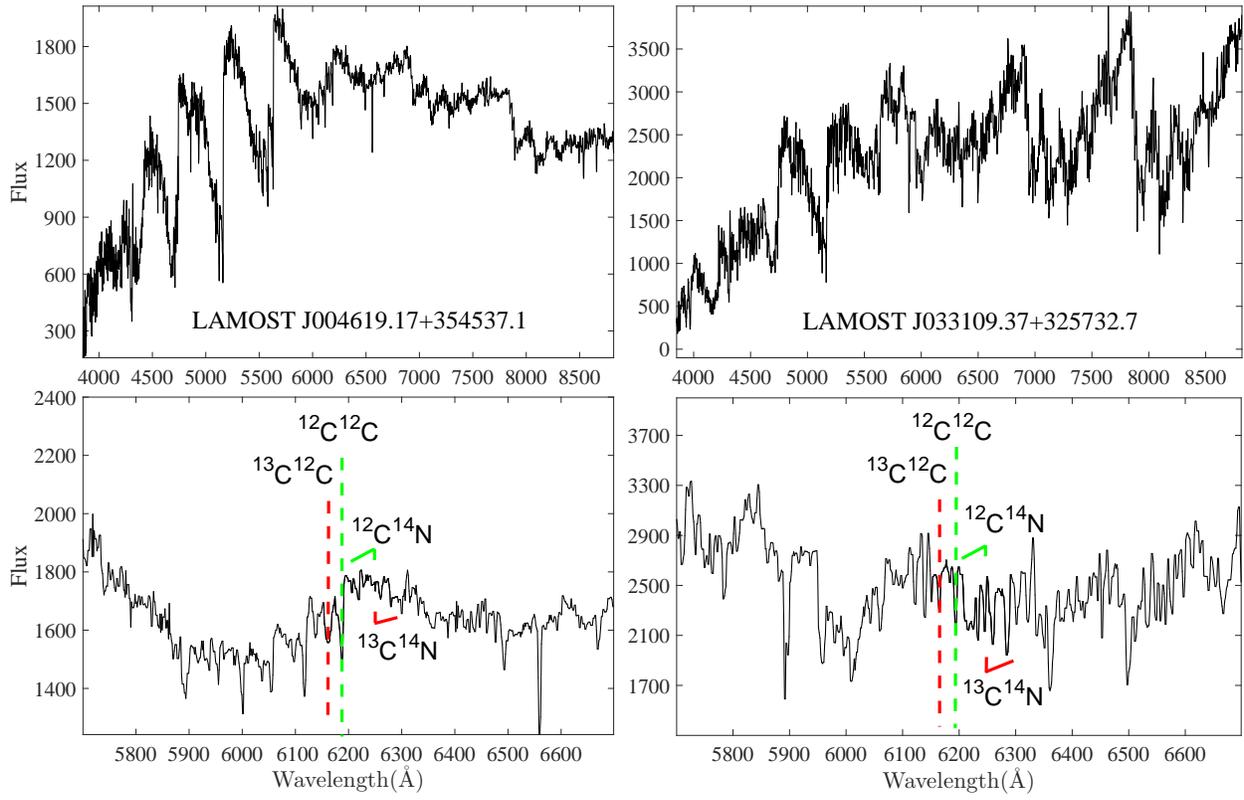}
\caption{Same as Fig. 8 for a C-J(H)- and a C-J(R)-type star. \label{cj_hr}}
\end{figure}

First, we manually select over 500 C-J star candidates with strong isotopic bands, then pick out 400 C-J stars with $j$ indexes greater than 4 as defined by \citet{demello2009}. Further, we divide the C-J stars into three types, i.e., C-J(N), C-J(H), and C-J(R) as described in \citet{lloyd1986}, which, respectively represent C-N, C-H, and C-R stars with intensity isotopic bands. The number of C-J(N), C-J(H), C-J(R) and C-J(UNKNOWN) stars among our 400 C-J stars is listed in Table~\ref{table:cj_type}, with C-J(UNKNOWN) representing stars which are unable to be classified as a certain type. From the classification result, we can clearly see that close to 90$\%$ of C-J stars are cool C-N type. Two example spectra of C-J(N) are plotted in Figure~\ref{cj_n}, and spectra of a C-J(H) and a C-J(R) star are shown in Figure~\ref{cj_hr}. The upper-left panel of Figure~\ref{cj_hr} is the spectrum of a C-J(H) star, i.e., LAMOST~J004619.17+354537.1; the upper right-panel is the spectrum of a C-J(R) star, i.e., J033109.37+325732.7.

\begin{figure}
\centering
\epsscale{1.0}
\plotone{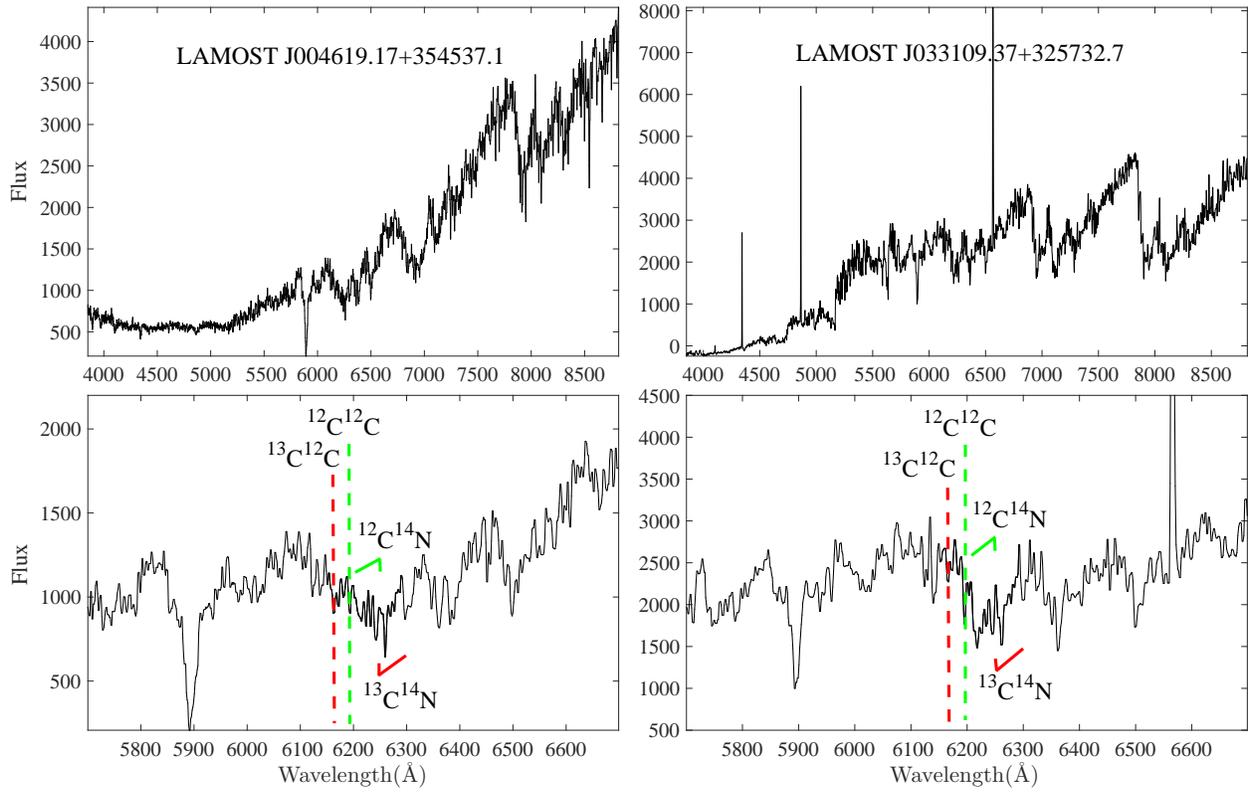}
\caption{Same as Fig. 8 for a C-J star with a composite spectrum and a C-J(N)-EM type star. \label{cj_binary_em}}
\end{figure}

In addition, we find 56 C-J stars with emission lines, and a cool C-J(N) star with composite spectra. Among these 56 emission-line (EM) stars, 55 are C-J(N) stars and one is a C-J(H) star. The spectrum with emission lines (J033109.37+325732.7) is shown in the upper-right panel, and the composite spectrum (J004619.17+354537.1) is plotted in the upper-left panel of Figure~\ref{cj_binary_em}.

In the bottom two panels of Figure~\ref{cj_n}--\ref{cj_binary_em}, we show the local spectra from 5700 to 6700 $\rm\AA$ of the upper two spectra, which are used to calculate the pseudo-continuum and the $j$ indexes. We also show the isotopic $^{13}$C$^{12}$C lines at 6168 $\rm\AA$ and the normal $^{12}$C$^{12}$C lines at 6192 $\rm\AA$ with red and green, respectively, and also the normal $^{12}$C$^{14}$N band at 6206 $\rm\AA$ and isotopic $^{13}$C$^{14}$N band at 6260 $\rm\AA$ with red and green colors. From these local spectra, we can see that the C-J stars show distinctly different C$_{2}$ and CN bands compared to the common C-H, C-R, and C-N stars.

\begin{figure}
\centering
\epsscale{1.0}
\plotone{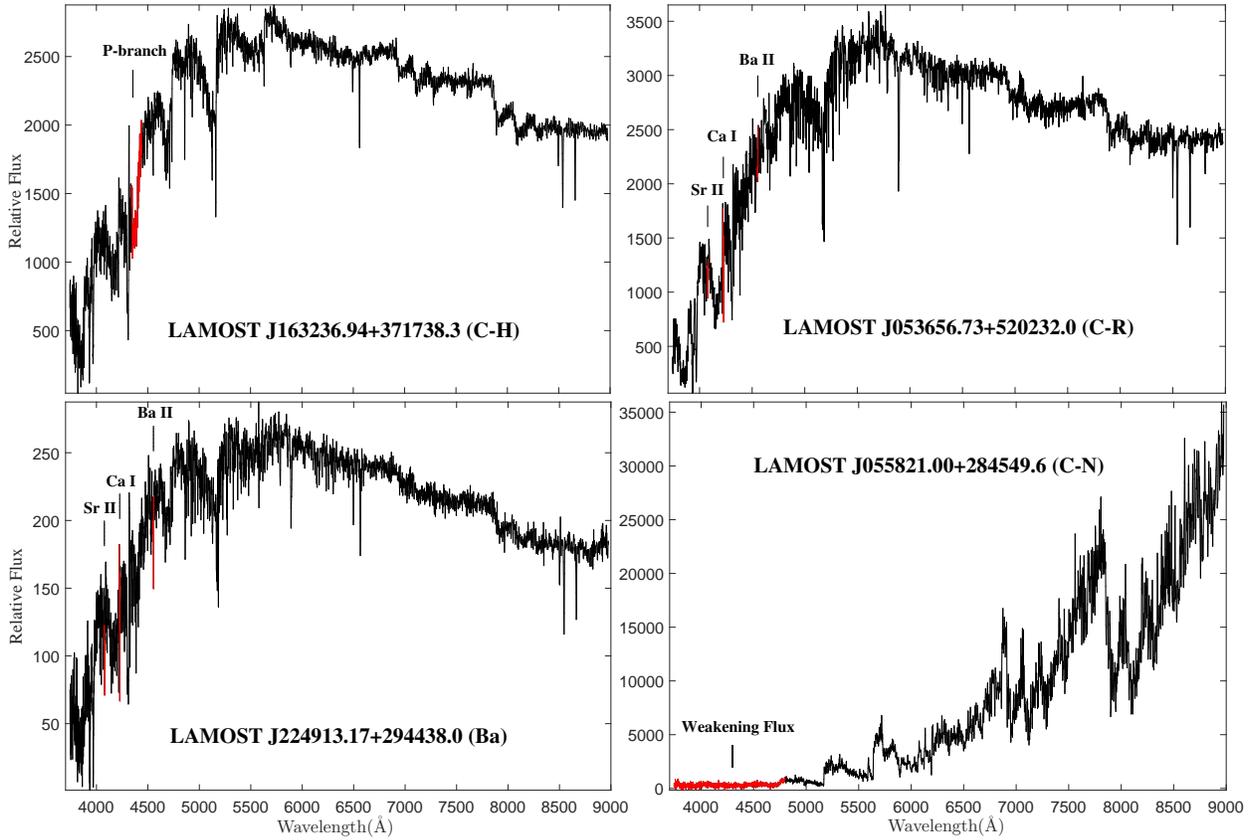}
\caption{The upper-left panel is the spectrum of a C-H star, and the red region is the P-branch band. The upper-right panel is the spectrum of a C-R star, and the red spectral lines are respectively Ba~II, Sr~II and
Ca I lines. The bottom left panel is the spectrum of a Barium star, and the Ba~II, Sr~II and Ca I lines are also marked by red color. The bottom right panel is the spectrum of a C-N star. \label{ba_ch_cr_cn}}
\end{figure}

 Aside from the 400 C-J stars, we identify 864 C-H stars, 226 C-R stars, 719 barium stars, 266 C-N stars, and 176 unclassified stars as listed in Table~\ref{table:spectra_type}. Among the 176 stars without spectral types, 12 of them are stars with composite spectra, and the other 164 stars have low spectral quality which make them unable to be classified. Four example spectra of the C-H, C-R, barium and C-N stars are displayed in Figure~\ref{ba_ch_cr_cn}, and their most prominent spectral lines, which are used to determine their spectral types, are marked in red.

\begin{figure}
\centering
\epsscale{1}
\plotone{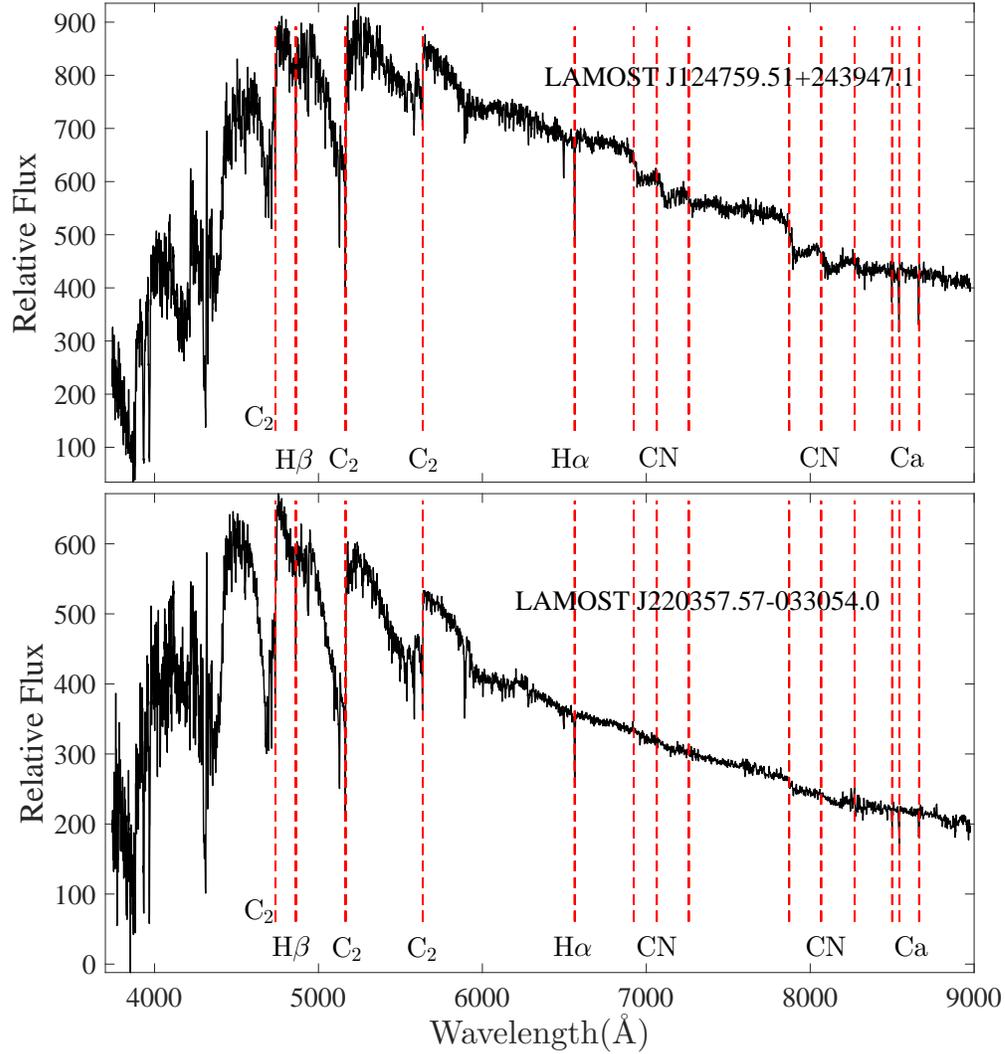}
\caption{Two spectra of G-type carbon stars, which remarkably show blue continua, strong C$_{2}$ bands, and strong narrow Balmer and Ca absorption lines. Sometimes, a fraction of G-type stars show weak CN band in the red part, for example, the object of J124759.51+243947.1. \label{spectra_Gtype}}
\end{figure}

\begin{figure}
\centering
\epsscale{1.0}
\plotone{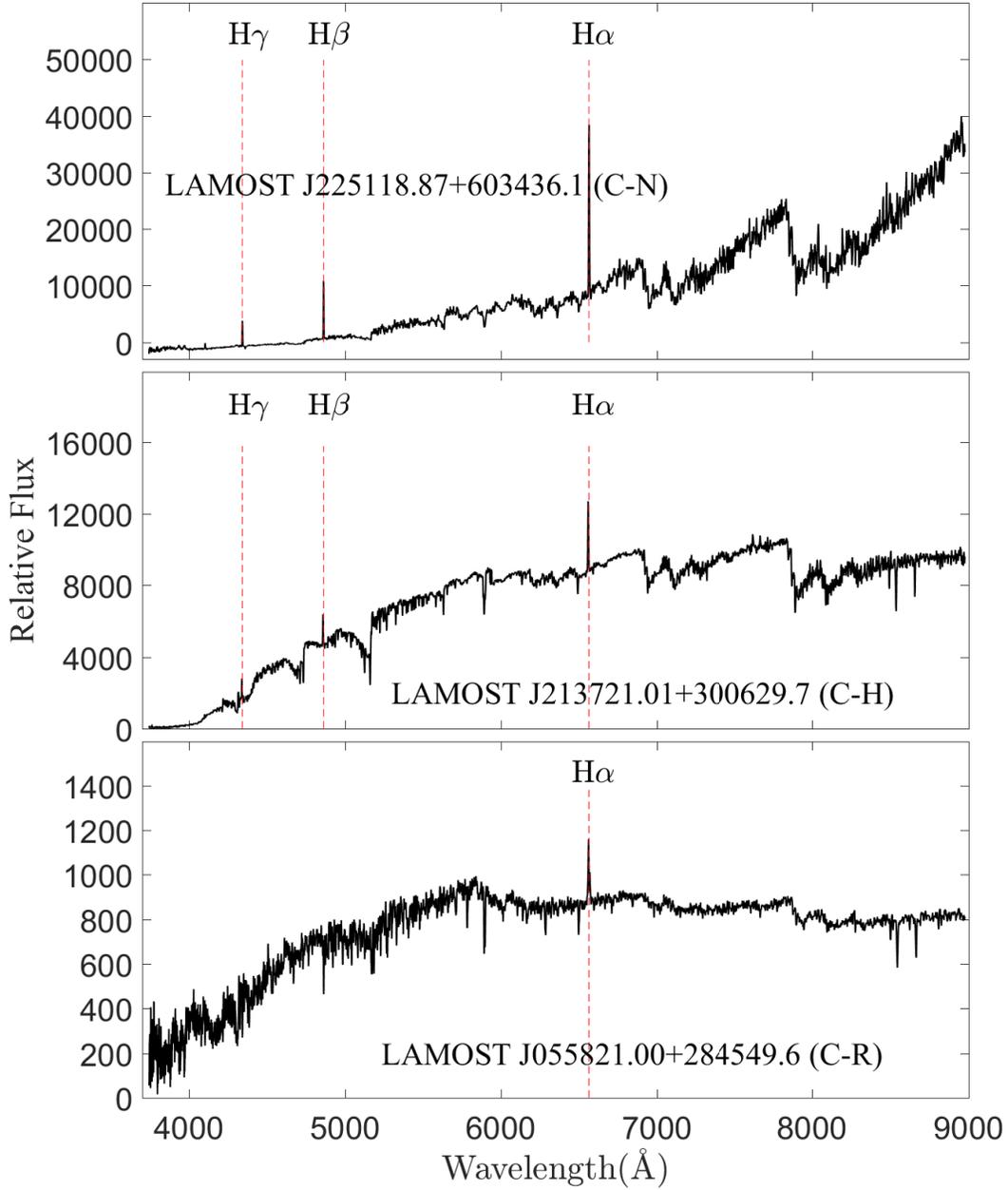}
\caption{Three spectra of EM stars: the upper panel shows a cool C-N type stars, the middle panel is a C-H type star, and the bottom one is a C-R type EM star. \label{spectra_em}}
\end{figure}

\begin{figure}
\centering
\epsscale{1.0}
\plotone{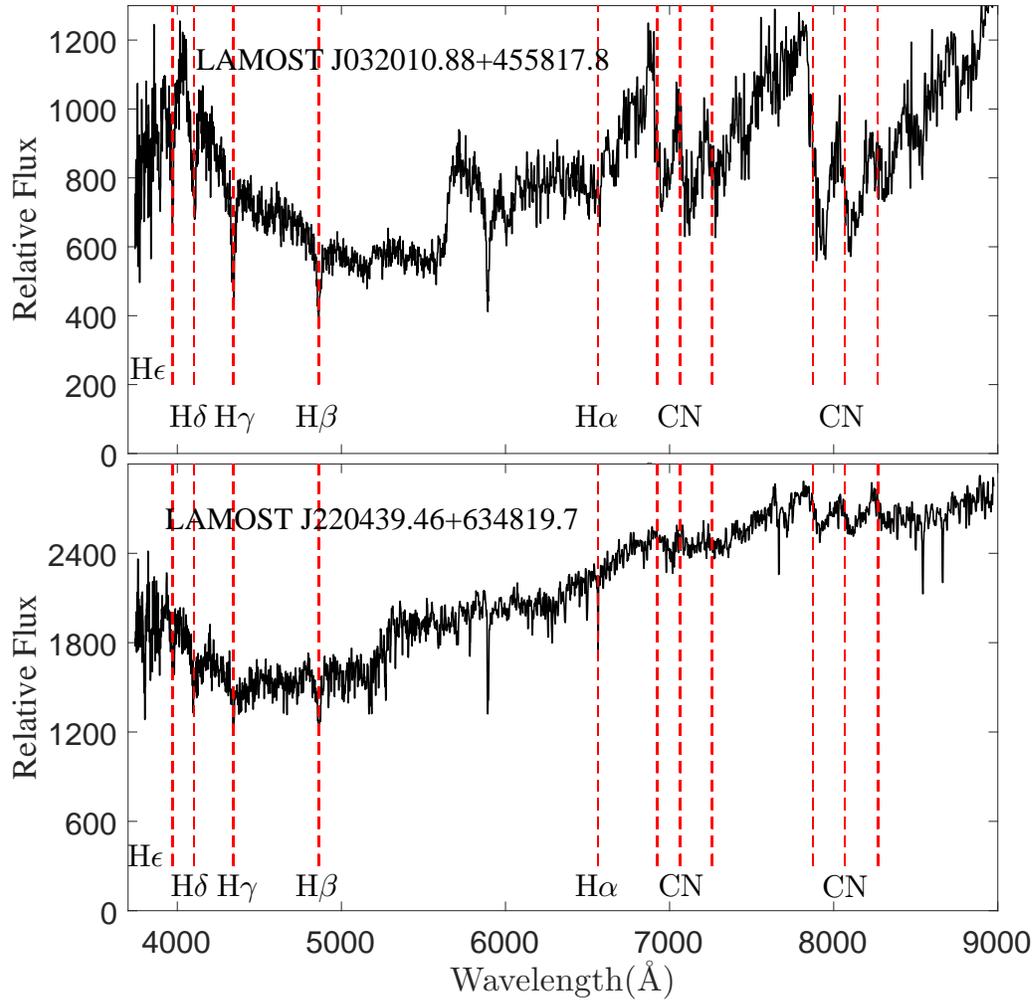}
\caption{Carbon stars with composite spectra, which display strong continua and the pressure-broadened hydrogen Balmer absorption lines of typical DA white dwarfs
in the blue part and clear C$_{2}$ bands and red CN bands in the red part. \label{spectra_binary}}
\end{figure}

\citet{green2013} found 134 G-type stars, 51 emission-line (EM) stars, and 9 stars with clear composite spectra, which have distinctly different spectral features from other carbon stars. According to the MK classification criteria, G-type stars can be classified as C-H stars, and most majority of EM stars are either C-N or C-J(N) stars. In our carbons stars, there are 308 G-type carbon stars, 93 EM stars, and 12 stars with composite spectra. Among the 308 G-type stars, 250 are new discoveries. We plot two example spectra in Figure~\ref{spectra_Gtype}. Among the 93 EM stars, 45 of them are reported for the first time. We plot three example spectra in Figure~\ref{spectra_em}, the upper panel is a cool C-N type star, the middle panel is a C-H type star, i.e., J213721.01+300629.7, and the bottom panel is a C-R type star, i.e., J055821.00+284549.6. The C-H and C-R type EM stars are extremely rare, and we only find two C-H type EM stars and one C-R type EM star in this paper. Of the 12 stars with composite spectra, seven of them are newly recognized in this work, and two example plots are shown in Figure~\ref{spectra_binary}.

The basic information on our carbon stars are listed in Table~\ref{table:basic_parameters_carbon}, which inclu des ``R.A.,'' ``Decl.,'' ``SNR$\_$r,'' ``SpType$\_$PL,'' ``G$\_$EM$\_$B,'' ``If$\_$new,'' and ``SpType$\_$MK.'' ``R.A.'' and ``Decl.'' are equatorial coordinates; ``SNR$\_$r'' is the $r$-band S/N;``SpType$\_$PL'' is the spectral type given by the LAMOST 1D pipeline; and ``G$\_$EM$\_$B'' has four values, which are respectively ``EM,'' ``G$\_$type,'' ``Binary,'' and ``NULL,'' and represent EM stars, G-type stars, stars with composite spectra, and other stars. ``If$\_$new,'' which has two values, ``NEW'' and ``NULL'', denotes whether it is a newly recognized star; and ``SpType$\_$MK'' gives the MK spectral classification results, which has 15 values. Among them, ``C-H,'' ``C-R,'' ``C-N,'' and ``Ba,'' respectively represent C-H, C-R, C-N, and barium stars, ``UNKNOWN'' means stars for which we cannot give a spectral type, and ``NULL'' is for stars which cannot be classified. For C-J stars, we set four values for them. ``C-J(H),'' ``C-J(R),'' ``C-J(N),'' and ``C-J(UNKNOWN),'' respectively represent C-H, C-R, C-N, and unknown stars with unusually strong isotopic carbon bands. For stars with emission lines, we use five values to represent them; these are ``C-H(EM),'' ``C-N(EM),'' ``C-R(EM),'' ``C-J(H)-EM,'' and ``C-J(N)-EM.'' Among these, ``C-H(EM),'' ``C-N(EM),'' and  ``C-R(EM),'' represent C-H, C-N, and C-R stars with emission lines, and ``C-J(H)-EM'' and ``C-J(N)-EM'' are C-H and C-N stars with strong isotopic carbon bands and emission lines in their spectra.

We match our carbon stars to the photometric catalogs of $GALEX$, Pan-STARRS, 2MASS, and $WISE$, and list their magnitude information in the machine-readable form of Table ~\ref{table:basic_parameters_carbon}. FUV and NUV are ultraviolet GALEX magnitudes; u, g, r, i, and z are five optical Pan-STARRS magnitudes; J, H, and K are three 2MASS magnitudes; W1, W2, and W3 are three WISE magnitudes, and ``...'' represents no photometry detection in this band. In addition, we match to the PPMXL, UCAC4, and TGAS \citet{gaia2016} catalogs, and list their proper motions and parallaxes in in the machine-readable form of Table~\ref{table:basic_parameters_carbon}. There, $\mu_{\alpha}\cos(\delta)_{\_ \rm P}$ and $\mu_{\delta}$$_{\_ \rm P}$ are proper motions in the R.A. and decl. directions from the PPMXL catalog, $\mu_{\alpha}\cos(\delta)_{\_ \rm U}$ and $\mu_{\delta}$$_{\_ \rm U}$ are from the UCAC4 catalog, $\mu_{\alpha}\cos(\delta)_{\_ \rm TG}$ and $\mu_{\delta}$$_{\_ \rm TG}$ are from the TGAS catalog, and parallax$_{\_ \rm TG}$ is parallax also given by the TGAS. Among our carbon stars, a total of 124 stars have astrometry parameters from the TGAS catalog, 43 of which have a parallax larger than 1 mas yr$^{-1}$, which shows that they have relatively reliable measurement values of the parallax.

\subsection{Spatial Distribution}

\begin{figure}
\centering
\epsscale{1.0}
\plotone{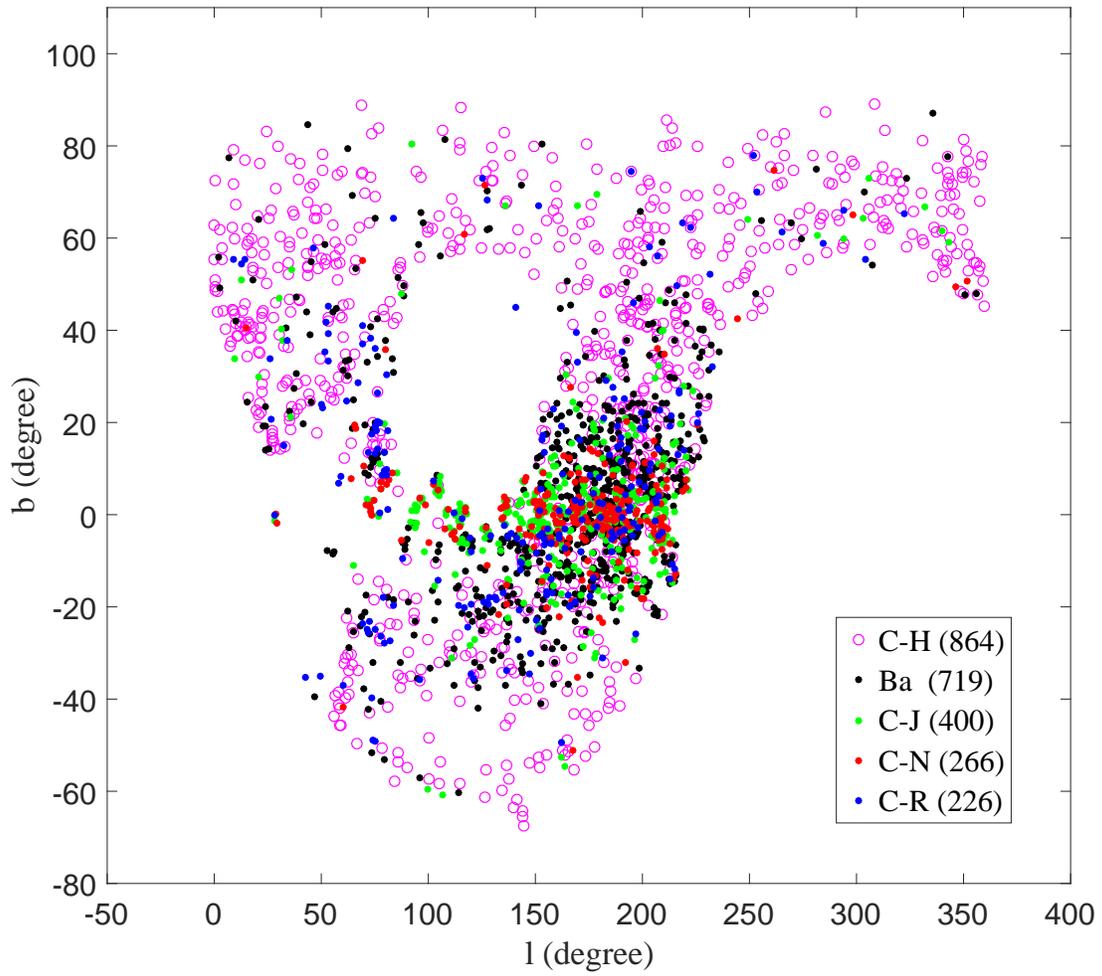}
\caption{Spatial distribution of the carbon stars reported in this paper. \label{gl_gb_distribution}}
\end{figure}

The spatial distribution in Galactic coordinates of the 2651 carbon stars  is plotted in Figure~\ref{gl_gb_distribution}, and colored symbols represent the types of carbon stars.  Among them, 12 C-N, 24 C-J, 568 C-H, 54 C-R and 125 barium stars are situated in the regions with $|b| >= 30^{\circ}$, and respectively account for about 4.5$\%$, 6$\%$, 65.7$\%$, 23.9$\%$, and 17.4$\%$ of the five types. As expected, the majority of C-N and C-J stars (95$\%$) and a large number of C-R and barium stars (75$\%$) are concentrated at low Galactic latitudes. A majority of the C-H stars (over 60$\%$) are located at the high latitudes, and the rest lie in the low latitudes.

\section{Photometry of carbon stars}

In this section, we investigate the 2651 carbon stars using photometric data from the ultraviolet to infrared. For the ultraviolet and optical bands, we respectively match with the GR6Plus7 catalog of the $GALEX$ \citealt{martin2005}) and the DR1 catalog of Pan-STARRS \citep{kaiser2002, kaiser2010, chambers2016} using the MAST CasJobs tool\footnote{http://galex.stsci.edu/casjobs/}.

$GALEX$\footnote{http://www.galex.caltech.edu/index.html} is a NASA small explorer mission launched in 2003 April, and it performed the first all-sky imaging and spectroscopic surveys in space in the ultraviolet band (1350-2750 $\rm\AA$). The main goal of GALEX is to investigate the causes and evolution of star formation in galaxies over the history of the universe in the ultraviolet band, making it is feasible to detect hot white dwarfs in unresolved binaries with main-sequence companions as early as G or K type, and cooler white dwarfs with early M-type or later companions as mentioned in \citet{green2013}.

Pan-STARRS\footnote{http://panstarrs.stsci.edu/} is a system for wide-field astronomical imaging developed and operated by the Institute for Astronomy at the University of Hawaii. Pan-STARRS1 (PS1) is the first part of Pan-STARRS to be completed and is the basis for Data Release 1 (DR1). It images the sky in the g, r, i, z, and y five broadband filters.

For the infrared bands, we match to 2MASS \citep{skrutskie2006} and WISE \citep{wright2010} using the X-match tool\footnote{http://cdsxmatch.u-strasbg.fr/xmatch} of SIMBAD astronomical database\footnote{http://simbad.u-strasbg.fr/simbad/}. 2MASS performed uniform precise photometry and astrometry in the near-infrared photometric bandpasses $J$ (1.25 $\rm \mu$m), $H$ (1.65 $\rm \mu$m), and $K$ (2.16 $\rm \mu$m) between 1997 June and 2001 February, and produced a point source catalog containing 470,992,970 sources and an extended source catalog of 1,647,599 sources covering 99.998$\%$ of the celestial sphere. $WISE$ began a mid-infrared survey of the entire sky in 2009 mid-December, and completed it in the year 2010. It mapped the whole sky in four infrared bands, $W1$, $W2$, $W3$, and $W4$, which were centered at 3.4, 4.6, 12, and 22 $\rm \mu$m using a 40 cm telescope feeding arrays with a total of 4 million pixels.

\subsection{GALEX Ultraviolet-detected stars}

$GALEX$ photometry is in two bands, far- and near-ultraviolet respectively (FUV: 1771-2831 $\rm\AA$ and NUV: 1344-1786 $\rm\AA$). The strong UV flux detected by GALEX in stellar systems can arise from a very hot blackbody, such as a hot white dwarf, or may also be associated with stellar activity \citep{green2013}.

\begin{figure}
\centering
\epsscale{1.0}
\plotone{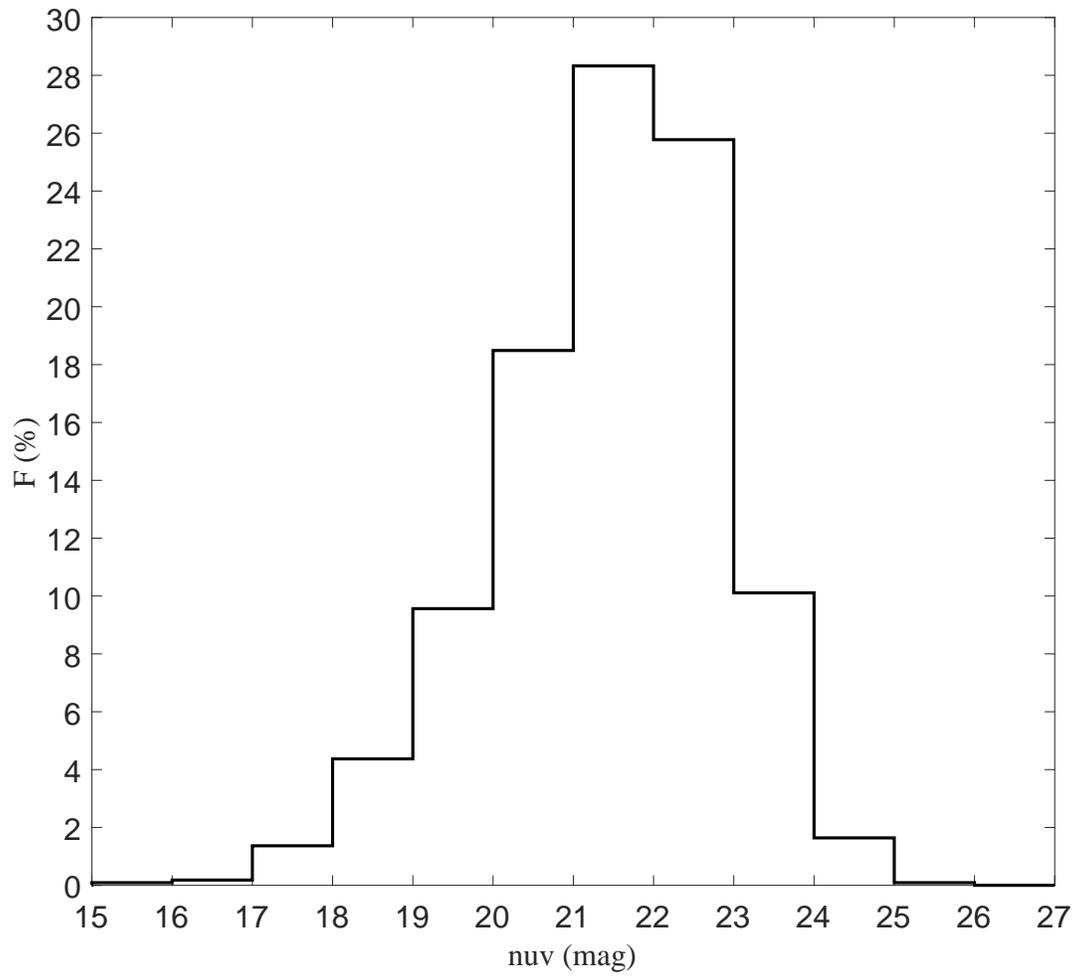}
\caption{Distribution of ultraviolet $GALEX$ NUV magnitudes. \label{nuv_distribution}}
\end{figure}

We match to the GR6Plus7 catalog of GALEX with a 6$^{''}$ search radius and use the ``fGetNearestObjEq'' function of CasJobs tool to only get the best matching result for each star. In all our carbon stars, we find a total of 1099 GALEX detections, and 1098 of them have NUV detections. We plot their NUV magnitude distribution in Figure~\ref{nuv_distribution}. It can be seen that the NUV magnitude is from 16 to 26 mag; over 90$\%$ of them lies in the range from 19 to 24 mag, and over 50$\%$ of them are in a narrow range from 21 to 23 mag.

Among our 308 G-type stars, about $81.2\%$ of them have GALEX magnitudes, which is an extremely high detection fraction as mentioned in \citet{green2013}. The number of C-H, C-R, C-J, C-N and barium stars having GALEX detections are listed in Table~\ref{table:spectra_type_GALEX}. From this table, it should be noted that there are a total of 991 hot C-H, C-R, and barium stars with GALEX detections, which represent 54.8$\%$ of all 1809 hot stars, and a total of 59 cool C-J and C-N stars with GALEX detections, which make up 8.9$\%$ of all 666 cool carbon stars. Thus, we can see that hot carbon stars have a higher $GALEX$ detection rate than cool stars as expected.

Further, we note that 1098 of the GALEX detections have NUV detections, and 25 of them have FUV detections. In these FUV-detected stars, one of them, J050736.14+305149.6, has a magnitude but no NUV detection, and its FUV magnitude error is over 0.5 mag. We list the FUV magnitudes, NUV magnitudes, and spectral types of these FUV detections in Table~\ref{table:GALEX}. From this table, we can clearly see that there are 11 C-H stars, two C-R stars, one C-J star, one C-N star, six barium stars, and five stars with unknown spectral type. Among the 11 hot C-H stars, three are them are G-type stars. Note that UV brightness can arise in young, active objects from their active regions, transition regions, or chromospheric emission. Among these 25 FUV detections, none of them have emission lines, and thus they are likely to be in binary systems with hot white dwarf companions.

\subsection{Pan-STARRS Optical-detected stars}

\begin{figure}
\centering
\epsscale{1.0}
\plotone{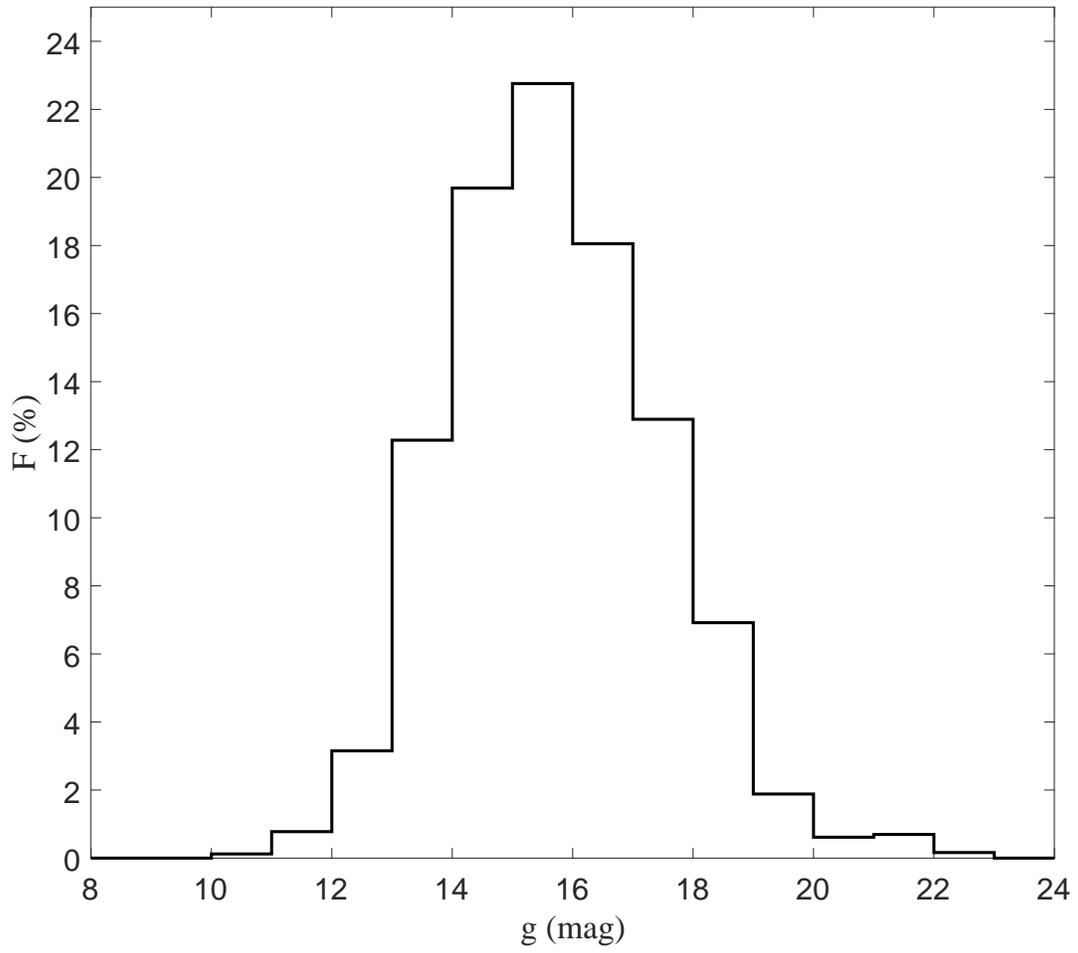}
\caption{Distribution of optical Pan-STARRS $g$ magnitudes. \label{g_distribution}}
\end{figure}

We match our carbon stars with the DR1 catalog of Pan-STARRS using a 3$^{''}$ search radius and also use the `fGetNearestObjEq' function of the CasJobs tool of Pan-STARRS to return the best matching result for each star. There are a total of 2608 stars having optical Pan-STARRS detections; the number of C-H, C-R, C-J, C-N, and barium stars having Pan-STARRS detections are listed in Table~\ref{table:spectra_type_GALEX}. We plot the $g$-band magnitude distribution of the 2608 Pan-STARRS detections in Figure~\ref{g_distribution}, and can clearly see that the magnitude range of the $g$ band is from 10 to 23 mag. In addition, over 90$\%$ of the carbon stars are located in the range from 13 to 19 mag, and over half of them are concentrated in the range between 14 and 17 mag.

\subsection{2MASS and WISE Infrared-detected stars}

We match all of our carbon star samples to the 2MASS catalog within 3$^{''}$, and find 2567 near-infrared 2MASS detections, dominating 96.8$\%$ of our carbon stars. Table~\ref{table:spectra_type_GALEX} lists the number of C-H, C-R, C-J, C-N, and barium stars that have 2MASS magnitudes.

We also match to the $WISE$ catalog within 3$^{''}$, and find 2550 mid-infrared $WISE$ detections, which means that 96.2$\%$ of our carbon stars have mid-infrared photometric magnitudes. Among these, there are 713 barium stars, 850 C-H stars, 222 C-R stars, 352 C-J stars, 241 C-N stars, and 161 carbon stars of unknown type, which are listed in Table~\ref{table:spectra_type_GALEX}. It should be noted that, when we match to 2MASS and $WISE$ catalogs, only results with distances nearest to the coordinates we upload are retained for stars with multiple results returned.

\begin{figure}
\centering
\epsscale{1.0}
\plotone{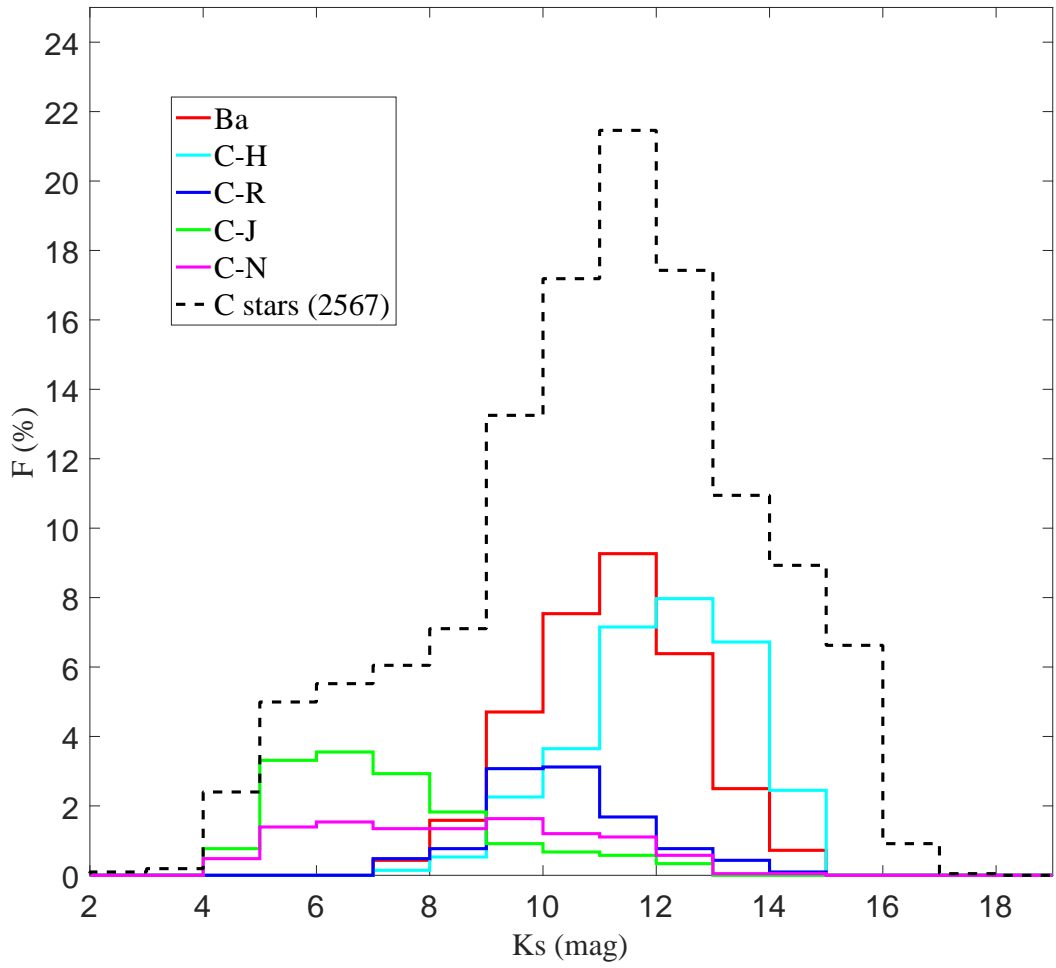}
\caption{Distribution of infrared 2MASS $K_{s}$ magnitudes. \label{ks_distribution}}
\end{figure}

In the infrared bands, we select the $Ks$ band for the investigation of the magnitude distribution. Considering a smaller extinction effect in the infrared bands, we plot the $Ks$ magnitude distributions of each type of carbon stars in Figure~\ref{ks_distribution}, together with the distribution of all 2MASS-detected stars. In this plot, the black dashed histogram shows the distribution of all 2567 stars; the red, cyan, blue, green, and magenta histograms respectively exhibit the distribution of the Ba, C-H, C-R, C-J and C-N stars. From the black dashed histogram, we can see that the $Ks$ magnitude of our stars is from 4 to 17 mag, and the peak of the distribution lies between 11 and 12 mag.  Further, from the five colored histograms, it can be seen that the cool C-J and C-N stars are the brightest, followed by the hot C-R, barium and C-H stars, as expected. Since the C-N and cool C-J stars are post-AGB stars with high luminosity, they can show very bright $Ks$ magnitude, while C-H and  barium stars are regarded as being in binary systems, and C-R stars were previously binaries, and thus these three types of stars have moderate distributions of magnitude.

\section{Summary}

In this paper, we retrieve carbon stars from the large spectral database of LAMOST DR4 and adopt an efficient machine-learning algorithm, i.e., the Bagging TopPush method.

As a supervised machine-learning method, positive samples are needed to train ranking models. We obtain 1050 positive samples selected from over 1600 spectra of SDSS and LAMOST carbon stars in the literatures, and cluster them into three groups. For each group, we respectively analyze the effect of spectral preprocessing methods on the algorithm performance and find that the so called ``nmap$+$pca'' method, which combines the ``nmap'' normalization (explained in appendix B) and the PCA dimension reduction has the highest performance for each group. The completeness and contamination for each group are also discussed.

We finally find 2651 carbon stars from LAMOST DR4, of which 1415 are reported for the first time. Among these stars, there are 308 G-type stars, 92 EM stars, and 12 spectral binaries, which show distinctly different spectral features from other stars. Among the 92 EM stars, we find two C-H type EM stars and one C-R type EM star, accounting for extremely the extremely low proportion of our emission-line carbon stars.

After identifying these carbons stars, we compare them with the classification results of the LAMOST 1D pipeline and two previous LAMOST catalogs \citep{sijianmin2015, jiwei2016}. At first, 63$\%$ of our carbon stars were correctly classified as carbon stars by the LAMOST 1D pipeline,  and the other 37$\%$ of our stars were recognized by our method. Among the 37$\%$ of stars misclassified, 74$\%$ were classified as G-type stars by the 1D pipeline, which suggests that it is necessary to construct new templates of carbon stars for the LAMOST 1D pipeline to improve the accuracy of the spectral classification. Then, a total of nine carbon stars present in previous catalogs were removed from our catalog because of low spectral quality. But, over 600 of our carbon stars were omitted from their catalogs at the same observation period, and we check and analyze the reasons why these stars in our catalog were excluded by them.

Based on a series of spectral features, we classify our carbon stars into five subtypes, i.e., C-H, C-R, C-J, C-N, and barium stars. We use $j$ indexes larger than 4 to find 400 C-J stars, and artificially classify the other stars using the spectral features summarized in Table~\ref{tab: classification_criteria}. Finally, we identify 864 C-H stars, 226 C-R stars, 400 C-J stars, 266 C-N stars, 719 barium stars, and 176 unclassified stars. The C-J stars are further divided into three subtypes of C-J(H), C-J(R), and C-J(N), and we find that close to 90$\%$ of them are cool C-J(N) stars.

We investigate the spatial distribution of the five types of carbon stars in the Galactic coordinates. For C-H stars, about over 60$\%$ of them are located at the region of high-latitude region, while others are located at the low latitude area. In addition, about 95$\%$ of cool C-N and C-J stars, and at least 75$\%$ of C-R and barium stars are concentrated in regions of low Galactic latitude.

 Aside from the 2651 carbon stars, we also find 17 CEMPTO stars, and preliminarily study their nature using atmospheric parameters obtained from low-resolution spectra. Up until now, there have only been about 20 CEMPTO stars were reported in the literatures; they are extremely rare stars, and high-resolution follow-up observations are needed for further identification.

At the end of this paper, we cross-match our carbons stars with the ultraviolet $GALEX$, optical Pan-STARRS, near-infrared 2MASS and mid-infrared $WISE$ catalogs, and study the magnitude distributions of the NUV, $g$, and $K_{s}$ bands. From the distribution of $K_{s}$, it can be seen that cool carbon stars are the brightest because of their post-AGB evolution stage, as expected.

In the future, it would be helpful to perform follow-up time domain photometric and high-resolution spectroscopic observations, which can be used to identify carbon stars and further investigate their nature.

\section{Acknowledgements}

We thank Bruce Margon, Wei Ji, Jian-Rong Shi, Wen-Yuan Cui, and Jincheng Guo for useful discussions. This work was supported by the National Natural Science Foundation of China (grant Nos. 11303036 and 11390371/4), the LAMOST FELLOWSHIP is supported by the Special Funding for Advanced Users, budgeted and administrated by the Center for Astronomical Mega-Science, Chinese Academy of Science (CAMS), and the National Basic Research Program of China (973 Program, 2014CB845700). Guoshoujing Telescope (the Large Sky Area Multi-Object Fiber Spectroscopic Telescope, LAMOST) is a National Major Scientific Project built by the Chinese Academy of Sciences. Funding for the project has been provided by the National Development and Reform Commission. LAMOST is operated and managed by the National Astronomical Observatories, Chinese Academy of Sciences.

\clearpage

\clearpage

\begin{figure}
\centering
\epsscale{1.0}
\plotone{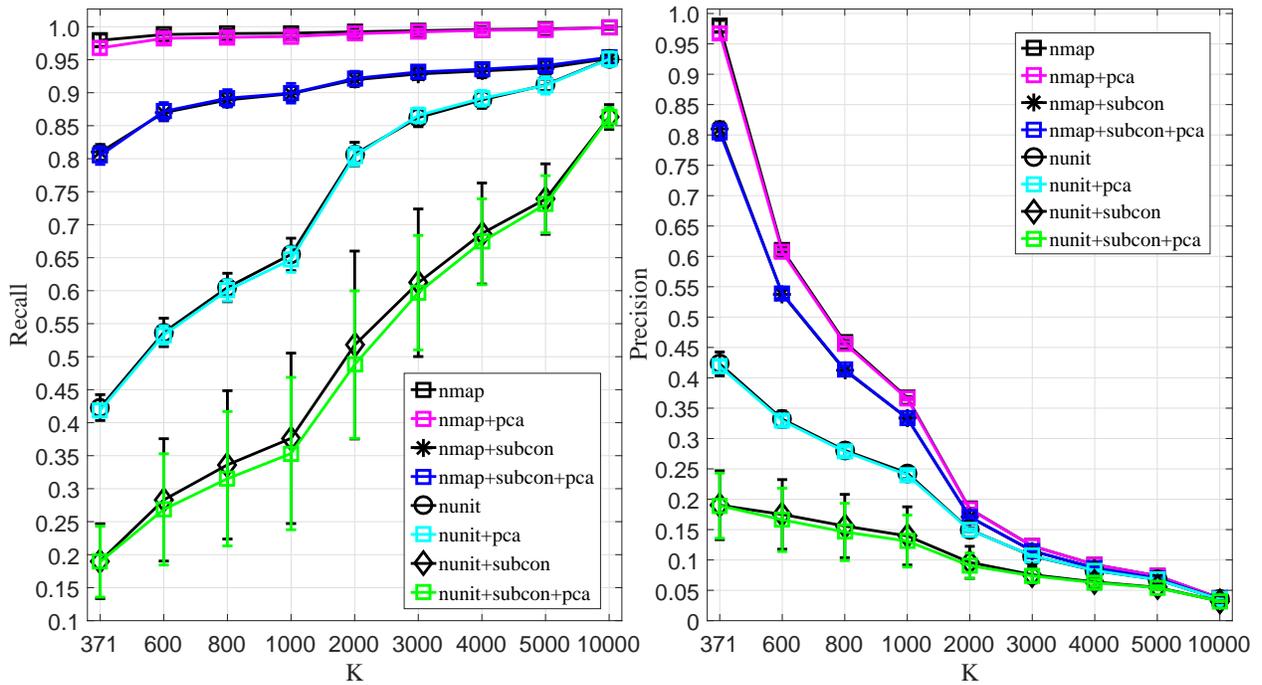}
\caption{The left panel is the relationship between recall and $K$ for positive samples of type I, and the right panel is the relationship between precision and $K$, where $K$ is the number of ranking results returned. The eight color lines in the two sub-plots represent the eight spectral preprocessing methods. \label{preprocess_I}}
\end{figure}

\clearpage

\begin{figure}
\centering
\epsscale{1.0}
\plotone{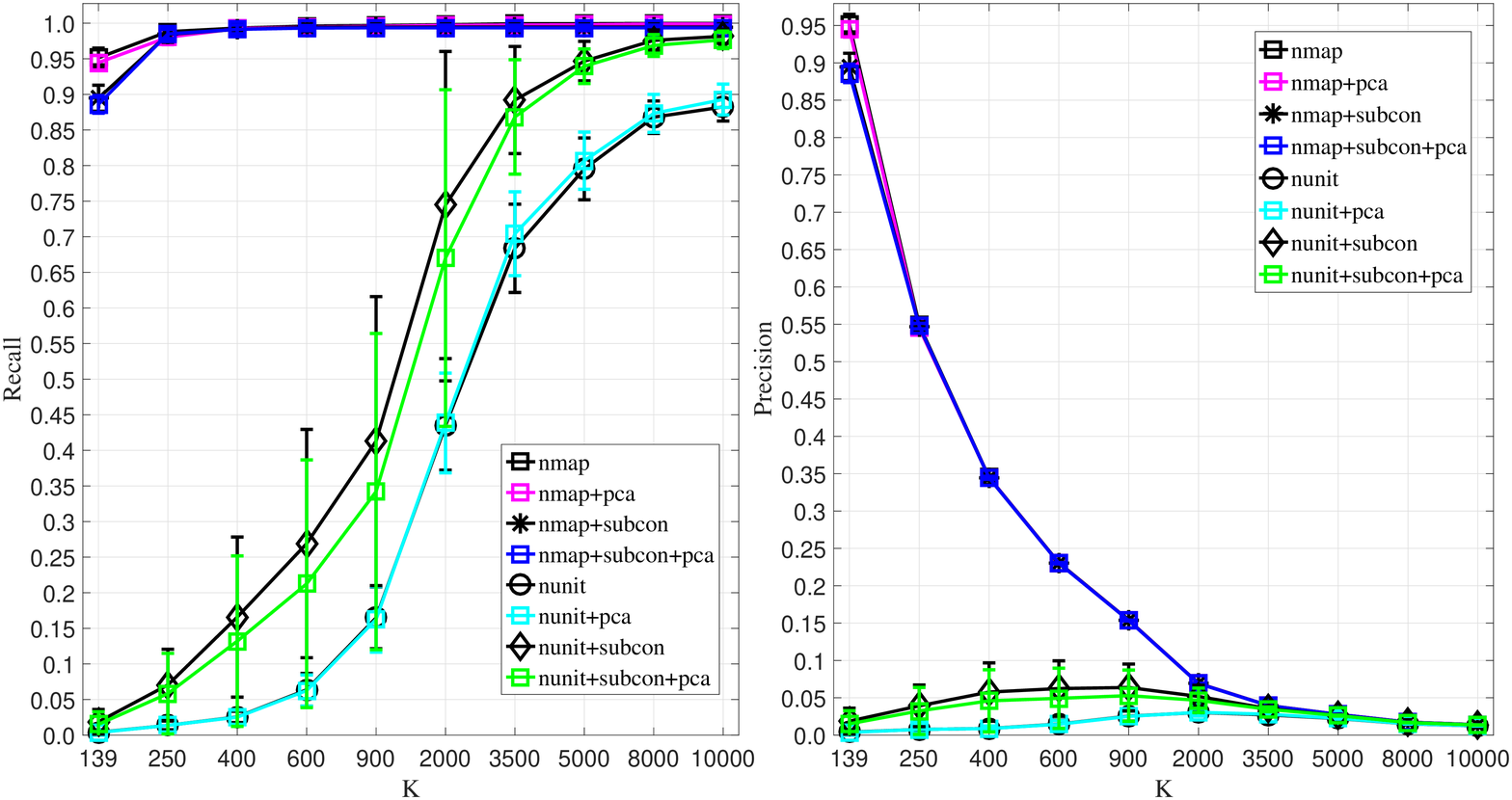}
\caption{Same as Fig.3 but for positive samples of type II. \label{preprocess_II}}
\end{figure}

\clearpage

\begin{figure}
\centering
\epsscale{1.0}
\plotone{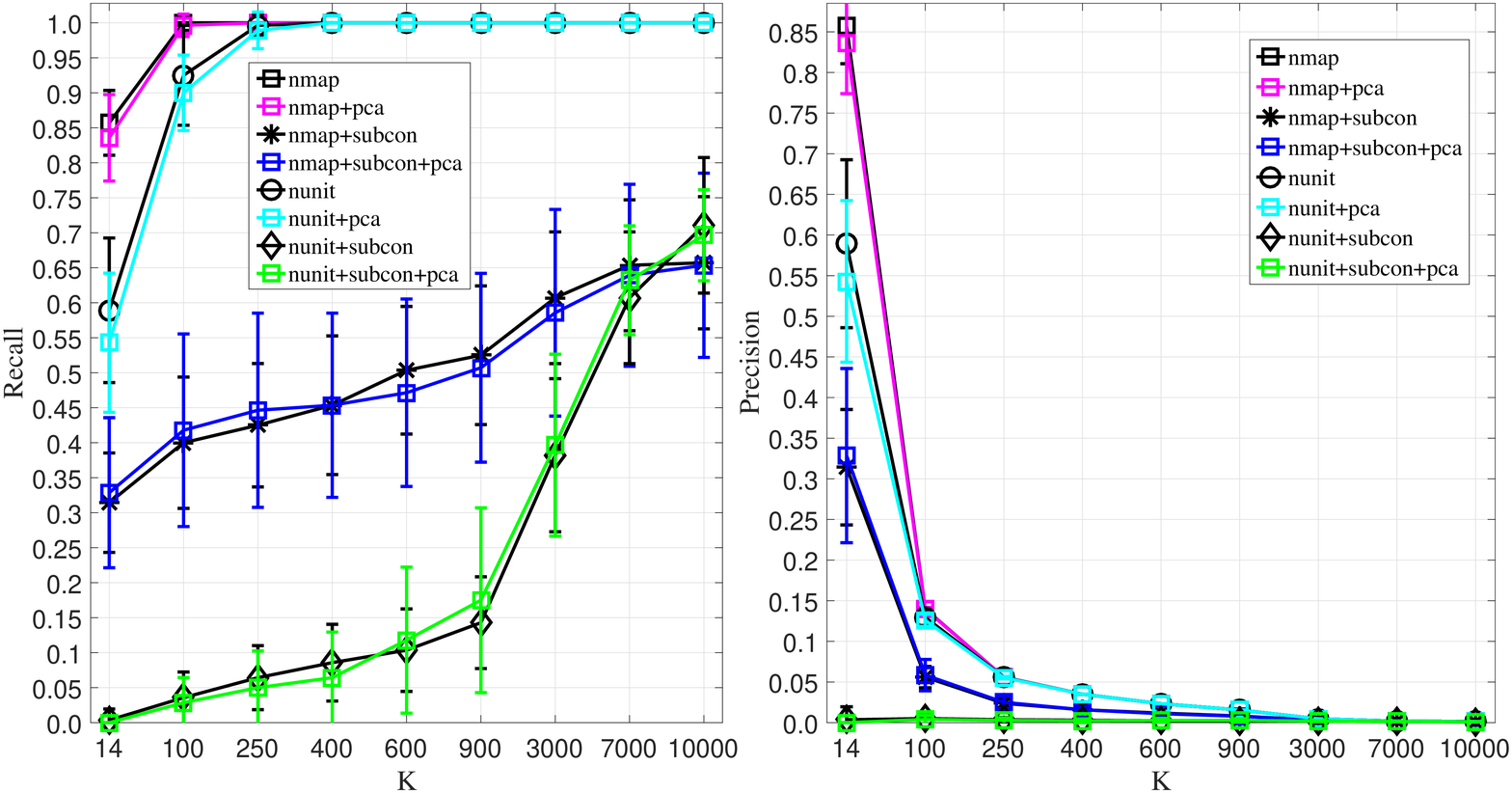}
\caption{Same as Fig.3 but for positive samples of type III. \label{preprocess_III}}
\end{figure}

\clearpage

\clearpage

\clearpage

\clearpage

\clearpage

\clearpage

\clearpage

\clearpage

\clearpage

\clearpage

\clearpage

\clearpage

\clearpage

\clearpage

\begin{table}
	\centering
	\caption{Classification Criteria of Carbon Stars.}\label{tab: classification_criteria}
	\begin{tabular}{c|l}
		\hline\hline
		Subtype & Criteria\\
		\hline
		\multirow{6}{*}{C-H} & (1) Prominence of the secondary P-branch head near 4342\,$\rm\AA$ \\
		& (2) Strong CH band \\
		& (3) H$\beta$ and Ba II at 4554 \,$\rm\AA$ are clearly noticeable \\
        & (4) H$\alpha$ and Ba II at 6496 \,$\rm\AA$ are noticeable \\
        & (5) Blend feature of Na I D1 and Na I D2 is not distinguishable\\
        & (6) Ca I at 4226 \,$\rm\AA$ is marginally noticeable \\
		\hline
		\multirow{4}{*}{C-R} & (1) Strong Ca I at 4226 \,$\rm\AA$ \\
        & (2) Na I D1 and Na I D2 blended lines have two distinct dips \\
        & (3) Weak H$\beta$ and Ba II at 4554 \,$\rm\AA$ blended with atomic and molecular lines \\
        & (4) Weak H$\alpha$ and Ba II at 6496 \,$\rm\AA$ blended with the CN bands around 6500 \,$\rm\AA$ \\
		\hline
		\multirow{3}{*}{C-N} & (1) No flux at $\lambda<4400$\,$\rm\AA$; some very late-type C-N can be flat even at $\lambda<5000$\,$\rm\AA$ \\
		& (2) Strong Ba II at 6496\,$\rm\AA$ \\
		& (3) Weak H$\alpha$ and isotopic C bands \\
        \hline
		\multirow{1}{*}{C-J} & (1) A high isotope ratio of $^{13}$C to $^{12}$C  with $j$ index $\geq$ 4 \\
        \hline
        \multirow{1}{*}{Ba} & (1) Strong lines of s-process elements, particularly Ba II at 4554 $\rm\AA$ and Sr II at 4077 $\rm\AA$ \\
		\hline\hline
	\end{tabular}
\end{table}

\clearpage

\begin{table*}
\begin{center}
\caption{Carbon Stars Reported in Previous Literature.}\label{table:previous_survey_1}
\begin{tabular}{ccccccccc}
\hline\hline
Literature & Total Number \tablenotemark{a} & if$\_$MK$\_$Sptype \tablenotemark{b}& C-H & C-R & Ba & C-J & C-N & Unknown\\
\hline
Alksnis et al. (2001) & 6891 & no &... & ... & ... & ... & ... & ... \\
Chrislieb et al. (2001) & 403 & no & ... & ... & ... & ... & ... & ...  \\
Margon et al. (2002) & 39 & no & ... & ... & ... & ... & ... & ...\\
Downes et al. (2004) & 251 & no & ... & ... & ... & ... & ... & ...  \\
Green (2013) & 1220 & no & ... & ... & ... & ... & ... & ... \\
Si et al. (2014) & 260 & no & ... & ...  & ... & ... & ... & ... \\
Si et al. (2015) & 183 & yes & 69 & 66 & ... & 4 & 33 & 10 \\
Ji et al. (2016) & 894 & yes & 339 & 259 & ... & ... & 108 & 82 \\
\hline\hline
\end{tabular}
\end{center}
\tablenotetext{a}{The total number of carbon stars reported in each literature.}
\tablenotetext{b}{This flag specifies whether spectral types were given using the MK classification system.}
\end{table*}

\clearpage

\begin{table*}
\begin{center}
\caption{The classification Results of Our carbon Stars given by the LAMOST 1D Pipeline.}\label{table:1D classification results}
\begin{tabular}{c|l}
		\hline\hline
		LAMOST Spectral Type & Number\\
		\hline
		\multirow{1}{*}{Carbon} & 1671\\
		\hline
        \multirow{1}{*}{Carbon white dwarf} & 2\\
        \hline
        \multirow{1}{*}{Unknown} & 149\\
        \hline
        \multirow{1}{*}{A9} & 1\\
        \hline
        \multirow{3}{*}{F} & F5: 1\\
        & F9: 3 \\
        & Total Number: 4 \\
        \hline
		\multirow{8}{*}{G} & G0: 6\\
        & G2: 8 \\
        & G5: 601 \\
        & G6: 6 \\
        & G7: 91 \\
        & G8: 12 \\
        & G9: 2 \\
        & Total Number: 726 \\
		\hline
		\multirow{8}{*}{K} & K0: 2 \\
        & K1: 23 \\
        & K2: 1 \\
        & K3: 22 \\
        & K4: 22 \\
        & K5: 23 \\
        & K7: 4 \\
        & Total Number: 97 \\
        \hline
        \multirow{1}{*}{M4} & 1\\
		\hline\hline
\end{tabular}
\end{center}
\end{table*}

\clearpage

\begin{table*}
\begin{center}
\caption{Comparison with Two Previous LAMOST Carbon Star Catalogs.}\label{table: si_ji}
\begin{tabular}{cccccccc}
\hline\hline
Literature & Total Number \tablenotemark{a} & Data Release \tablenotemark{b} & $N\_$1\tablenotemark{c} & $N\_$2 \tablenotemark{d}& $N\_$test \tablenotemark{e} & $N\_$version \tablenotemark{f} & $N\_$algorithm \tablenotemark{g}\\
\hline
Si et al. (2015) & 183 & Pilot & 4 & 33 & 4 & 12 & 17 \\
Ji et al. (2016) & 894 & DR2 & 5 & 575 & 122 & 50 & 403  \\
\hline\hline
\end{tabular}
\end{center}
\tablenotetext{a}{The total number of carbon stars reported in each publication.}
\tablenotetext{b}{The LAMOST data version used in each publication.}
\tablenotetext{c}{The number of carbon stars included in the publication but excluded from our catalog.}
\tablenotetext{d}{The number of carbon stars excluded from the publication but included in our catalog.}
\tablenotetext{e}{Among the N$\_$2 carbon stars excluded from the publication, the number of carbon stars observed in the test nights.}
\tablenotetext{f}{Among the N$\_$2 carbon stars excluded from the publication, the number of carbon stars excluded from the old version of the data set used in the publication.}
\tablenotetext{g}{Among the N$\_$2 carbon stars excluded from the publication, the numbers of carbon stars removed by the selection method in the publication.}
\end{table*}

\clearpage

\begin{table*}[htbp]
	\centering
	\caption{The Number of Our Carbon Stars Excluded by Each Step of the Selection Criteria of \citet{jiwei2016}.}\label{tab:steps}
	\begin{tabular}{l|c|cc}
		\hline\hline
	    Step & Criteria &  \multicolumn{2}{c}{Number of Candidates Removed} \\
		\hline
		1 & S/N$(i)>10$ and leave only one epoch for multiply observed stars &  \multicolumn{2}{c}{32} \\
		\hline		
        2 & No radial velocity  &  \multicolumn{2}{c}{25} \\
		\hline
		3 & Equations (2)--(4) &  \multicolumn{2}{c}{138} \\
		\hline
        \multirow{2}{*}{4}& CN7065\,$\ge4$ or CN7820\,$\ge8$  & &\multirow{2}{*}{113}  \\
		& CN7065\,$<4$ and CN7820\,$<8$ with S/N$(g)>20$ and S/N$(i)>20$   &  \\
		\hline
		5 & CN7065\,$\ge2$ and \ctwo\,$\ge-13$ & \multicolumn{2}{c}{54}  \\
		\hline
		6 & $K_s<14.5$ and $J-K_s>0.45$ &  \multicolumn{2}{c}{41}  \\
		\hline\hline
	\end{tabular}
\end{table*}

\clearpage

\begin{landscape}
\begin{table*}
\begin{center}
\caption{Equatorial Coordinates, S/Ns, Atmospheric Parameters, and Spectral Types of the 17 CEMP Turnoff star Candidates.}\label{table:basic_parameters_cempto}
\begin{tabular}{cccccccc}
\hline\hline
Designation & R.A. & Decl. & S/N$\_$r & Teff & log(g) & [Fe/H] & SpType$\_{\rm PL}$ \\
 & (degree) & (degree) &  & (K) &  &  & \\
\hline
J012514.16+352233.0 & 21.309014 & 35.375843 & 31 & 5982$\pm$44 & 3.87$\pm$0.07 & -2.05$\pm$0.07 & F2 \\
J025723.25+331638.6 & 44.346896 & 33.277406 & 32 & 5624$\pm$41 & 3.90$\pm$0.08 & -1.76$\pm$0.06 & G0 \\
J074637.32+291941.8 & 116.655510 & 29.328284 & 31 & 5858$\pm$45 & 3.90$\pm$0.09 & -2.04$\pm$0.07 & F0 \\
J082459.32+302542.0 & 126.247187 & 30.428351 & 43 & 6068$\pm$40 & 4.02$\pm$0.06 & -2.19$\pm$0.06 & A7 \\
J091243.72+021623.4 & 138.182182 & 2.273190 & 53 & 6001$\pm$36 & 3.91$\pm$0.06 & -2.17$\pm$0.05 & F2 \\
J093539.71+283138.7 & 143.915485 & 28.527433 &¡¡40 & 5847$\pm$40 & 3.74$\pm$0.07 & -2.24$\pm$0.05 & F4 \\
J095406.90+063728.2 & 148.528758 & 6.624504 & 48 & 5943$\pm$37 & 4.04$\pm$0.05 & -1.88$\pm$0.05 & F0 \\
J112535.49+414242.5 & 171.397877 & 41.711829 & 62 & 5997$\pm$41 & 4.00$\pm$0.06 & -1.96$\pm$0.06 & F0 \\
J121106.02+321044.8 & 182.775099 & 32.179121 & 25 & 5831$\pm$50 & 3.73$\pm$0.09& -2.14$\pm$0.07 & F5 \\
J121244.80+315134.0 & 183.186669 & 31.859463 & 69 & 5919$\pm$21 & 3.68$\pm$0.04 & -2.28$\pm$0.03 & F5 \\
J121817.74+295300.4 & 184.573949 & 29.883457 & 59 & 5861$\pm$25 & 4.01$\pm$0.04 & -1.64$\pm$0.03 & F5 \\
J125725.33+335356.8 & 194.355549 & 33.899133 & 100 & 5866$\pm$23 & 3.92$\pm$0.04 & -1.34$\pm$0.03 & F2 \\
J133539.75+152359.8 & 203.915639 & 15.399948 & 42 & 5768$\pm$43 & 3.66$\pm$0.08 & -2.23$\pm$0.06 & F5 \\
J140753.27+473517.3 & 211.971998 & 47.588155 & 58 & 5913$\pm$34 & 3.98$\pm$0.06 & -1.62$\pm$0.05 & F5 \\
J144427.16+314601.9 & 221.113201 & 31.767221 & 23 & 5698$\pm$45 & 3.31$\pm$0.1 & -2.42$\pm$0.06 & G3 \\
J151139.17+335252.9 & 227.913238 & 33.881386 & 102 & 5920$\pm$25 & 3.68$\pm$0.04 & -2.26$\pm$0.03 & F2 \\
J163008.22+231438.7 & 247.534280 & 23.244110 & 160 & 5850$\pm$22 & 3.90$\pm$0.03 & -1.90$\pm$0.03 & F5 \\
\hline\hline
\end{tabular}
\end{center}
\end{table*}
\end{landscape}

\clearpage

\begin{landscape}
\begin{table*}
\begin{center}
\caption{Ultraviolet, Optical, and Infrared Magnitudes of 17 CEMP Turnoff Star Candidates.}\label{table:magnitudes_cempto}
\begin{tabular}{cccccccccccccc}
\hline\hline
Designation & fuv & nuv & $u$ & $g$ & $r$ & $i$ & $z$ & $J$ & $H$ & $K$ & $W1$ & $W2$ & $W3$ \\
 & (mag) & (mag) & (mag) & (mag) & (mag) & (mag) & (mag) & (mag) & (mag) & (mag) & (mag) & (mag) & (mag) \\
\hline
J012514.16+352233.0 & ... $^{a}$& 19.87 & $...$ & $...$ & $...$ & $...$ & $...$ & 15.99 & 15.71 & 15.74 & 15.53 & 15.71 & 12.88 \\
J025723.25+331638.6 & ... & 20.53 & 18.07 & 17.07 & 16.63 & 16.44 & 16.36 & 15.52 & 15.08 & 14.91 & 14.91 & 14.99 & 12.47 \\
J074637.32+291941.8 & ... & 19.14 & 17.30 & 16.31 & 16.03 & 15.91 & 15.89 & 15.10 & 14.88 & 14.98 & 14.74 & 14.70 & 11.82 \\
J082459.32+302542.0 & ... & 18.89 & 17.15 & 16.24 & 15.94 & 15.85 & 15.82 & 15.10 & 14.64 & 14.81 & 14.71 & 14.76 & 11.82 \\
J091243.72+021623.4 & ... & ... & 16.55 & 15.67 & 15.37 & 15.24 & 15.21 & 14.47 & 14.18 & 14.07 & 14.03 & 14.03 & 11.78 \\
J093539.71+283138.7 & ... & 18.14 & 16.23 & 15.39 & 15.06 & 14.94 & 14.90 & 14.15 & 13.76 & 13.76 & 13.76 & 13.77 & 12.63 \\
J095406.90+063728.2 & ... & 19.58 & 17.72 & 16.76 & 16.43 & 16.32 & 16.27 & 15.42 & 15.17 & 15.24 & 15.18 & 15.16 & 12.62 \\
J112535.49+414242.5 & ... & 16.89 & 14.97 & 14.51 & 14.63 & 14.98 & 13.65 & 12.86 & 12.57 & 12.53 & 12.51 & 12.53 & 11.97 \\
J121106.02+321044.8 & ... & 20.15 & 18.47 & 17.57 & 17.27 & 17.13 & 17.11 & 16.22 & 15.72 & 15.67 & 16.00 & 16.08 & 12.50 \\
J121244.80+315134.0 & ... & 16.13 & 14.53 & 13.67 & 13.35 & 14.46 & 13.25 & 12.44 & 12.14 & 12.15 & 12.09 & 12.13 & 12.65 \\
J121817.74+295300.4 & ... & 17.27 & 15.32 & 14.44 & 14.21 & 14.01 & 14.03 & 13.18 & 12.95 & 12.84 & 12.85 & 12.85 & 12.42 \\
J125725.33+335356.8 & ... & 18.25 & 16.96 & 15.98 & 15.66 & 15.55 & 15.56 & 14.75 & 14.52 & 14.43 & 14.38 & 14.41 & 12.41 \\
J133539.75+152359.8 & ... & 18.50 & 16.45 & 15.57 & 15.20 & 15.08 & 15.03 & 14.21 & 13.87 & 13.88 & 13.84 & 13.88 & 12.70 \\
J140753.27+473517.3 & ... & 17.50 & 15.59 & 14.59 & 14.38 & 14.14 & 14.15 & 13.31 & 13.03 & 12.99 & 12.89 & 12.88 & 12.53 \\
J144427.16+314601.9 & ... & 18.92 & 17.01 & 16.03 & 15.64 & 15.55 & 15.50 & 14.71 & 14.29 & 14.31 & 14.28 & 14.34 & 12.86 \\
J151139.17+335252.9 & ... & 16.70 & 14.89 & 13.96 & 13.68 & 13.55 & 13.53 & 12.74 & 12.47 & 12.43 & 12.39 & 12.40 & 12.65 \\
J163008.22+231438.7 & ... & 17.35 & 15.41 & 14.50 & 14.34 & 14.02 & 14.01 & 13.20 & 12.86 & 12.81 & 12.80 & 12.82 & 12.09 \\
\hline\hline
\end{tabular}
\end{center}
\tablenotetext{a}{``...'' means no value is available.}
\end{table*}
\end{landscape}

\clearpage

\begin{landscape}
\begin{table*}
\begin{center}
\caption{Radial Velocities and Proper Motions of 17 CEMP Turnoff Star Candidates.}\label{table:motion_cempto}
\begin{tabular}{cccccc}
\hline\hline
Designation & RV & $\mu_{\alpha}\cos(\delta)_{\_ \rm P}$ & $\mu_{\delta}$$_{\_ \rm P}$ & $\mu_{\alpha}\cos(\delta)_{\_ \rm U}$ & $\mu_{\delta}$$_{\_ \rm U}$ \\
 & (km s$^{-1}$) & (mas yr$^{-1}$) & (mas yr$^{-1}$) & (mas yr$^{-1}$) & (mas yr$^{-1}$)  \\
\hline
J012514.16+352233.0 & -263$\pm$4 & 2.7$\pm$5.0 & -3.8$\pm$5.0 & ...$^{a}$ & ... \\
J025723.25+331638.6 & -235$\pm$3 & 8.6$\pm$4.3 & -32.2$\pm$4.3 & 5.0$\pm$3.8 & -32.1$\pm$3.7 \\
J074637.32+291941.8 & 78$\pm$4 & 9.0$\pm$4.1 & 0.3$\pm$4.1 & 10.8$\pm$5.5 & -10.5$\pm$6.1 \\
J082459.32+302542.0 & 24$\pm$4 & 4.3$\pm$4.3 & -32.0$\pm$4.3 & 8.9$\pm$4.0 & -28.5$\pm$4.5 \\
J091243.72+021623.4 & 115$\pm$4 & -9.5$\pm$3.6 & -29.6$\pm$3.6 & -1.9$\pm$2.2 & -30.3$\pm$2.4 \\
J093539.71+283138.7 & -79$\pm$3 & 53.6$\pm$3.7 & -73.0$\pm$3.7 & 55.0$\pm$2.7 & -76.1$\pm$3.0 \\
J095406.90+063728.2 & 283$\pm$3 & -13.5$\pm$4.1 & -21.3$\pm$4.1 & -15.6$\pm$18 & -15.9$\pm$18.1 \\
J112535.49+414242.5 & -288$\pm$3 & -23.8$\pm$4.0 & -38.3$\pm$4.0 & -23.0$\pm$1.5 & -39.6$\pm$1.8 \\
J121106.02+321044.8 & 45$\pm$4 & 8.8$\pm$4.8 & -13.6$\pm$4.8 & ... & ... \\
J121244.80+315134.0 & 29$\pm$2 & -61.6$\pm$4.0 & -73.3$\pm$4.0 & -63.3$\pm$1.2 & -75.5$\pm$1.4 \\
J121817.74+295300.4 & -62$\pm$2 & -41.7$\pm$5.0 & -38.3$\pm$5.0 & -38.2$\pm$2.0 & -33.2$\pm$2.7 \\
J125725.33+335356.8 & 36$\pm$1 & -31.8$\pm$3.8 & -16.4$\pm$3.8 & -27.0$\pm$2.3 & -17.9$\pm$2.8 \\
J133539.75+152359.8 & 143$\pm$4 & -32.4$\pm$4.1 & -59.9$\pm$4.1 & -30.0$\pm$3.1 & -61.6$\pm$3.7\\
J140753.27+473517.3 & 23$\pm$2 & -25.1$\pm$3.6 & -15.5$\pm$3.6 & -22.5$\pm$1.5 & -14.0$\pm$1.6 \\
J144427.16+314601.9 & -103$\pm$3 & -22.7$\pm$4.2 & -10.4$\pm$4.2 & -25.4$\pm$2.9 & -3.3$\pm$3.3 \\
J151139.17+335252.9 & -337$\pm$2 & -54.5$\pm$4.3 & -70.8$\pm$4.3 & -48.7$\pm$2.8 & -72.1$\pm$2.4 \\
J163008.22+231438.7 & -244$\pm$2 & -31.5$\pm$4.0 & 56.0$\pm$4.0 & -25.6$\pm$3 & 54.5$\pm$3.4 \\
\hline\hline
\end{tabular}
\end{center}
\end{table*}
\end{landscape}

\clearpage

\begin{table*}
\begin{center}
\caption{Number of Each type of C-J Stars.}\label{table:cj_type}
\begin{tabular}{ccccc}
\hline\hline
C-J$\_{\rm TN}\tablenotemark{a} $ & C-J(H) & C-J(R) & C-J(N) & C-J(UNKNOWN) \\
\hline
400 & 41 (10.25$\%$) & 2 (0.50$\%$) & 353 (88.25$\%$) & 4 (1.00$\%$) \\
\hline\hline
\end{tabular}
\end{center}
\tablenotetext{a}{The total number of our C-J stars.}
\end{table*}

\clearpage

\begin{table*}
\begin{center}
\caption{number of Each Type of Carbon Stars.}\label{table:spectra_type}
\begin{tabular}{ccccccc}
\hline\hline
C-TN \tablenotemark{a} & C-H & C-R & C-J & C-N & Barium & Unknown\\
\hline
2651 & 864 & 226 & 400 & 266 & 719 & 176\\
\hline\hline
\tablenotetext{a}{C-TN is the total number of our carbon stars.}
\end{tabular}
\end{center}
\end{table*}

\clearpage

\begin{landscape}
\begin{table*}
\begin{center}
\caption{Basic Parameters of the Carbon Stars Reported in This Paper.}\label{table:basic_parameters_carbon}
\begin{tabular}{cccccccc}
\hline\hline
Designation & R.A. & Decl. & S/N$\_$r & SpType$\_$PL & G$\_$EM$\_$B & If$\_$new & SpType$\_$MK \\
 & (degree) & (degree) &  &  &  &  & \\
\hline
J085222.34+494610.3 & 133.093100 & 49.769550 & 458 & K4 & NULL & NULL & C-R \\
J004619.17+354537.1 & 11.579886 & 35.760306 & 31 & Carbon & NULL & NULL & C-J(H) \\
J005917.52+315605.4 & 14.823016 & 31.934838 & 130 & Carbon & EM & new & C-H(EM) \\
J164406.42+470635.7 & 251.026756 & 47.109933 & 108 & G5 & NULL & NULL & Ba \\
J101754.70+251201.0 & 154.477944 & 25.200283 & 112 & Carbon & G$\_$type & NULL & C-H \\
J081747.79+290531.8 & 124.449140 & 29.092172 & 14 & Non & NULL & NULL & UNKNOWN \\
J200607.53+460847.2 & 301.531408 & 46.146456 & 216 & Carbon & NULL & NULL & C-N \\
J033109.37+325732.7 & 52.789053 & 32.959109 & 17 & Carbon & NULL & new & C-J(R) \\
J055821.00+284549.6 & 89.587501 & 28.763803 & 62 & G5 & EM & NULL & C-R(EM) \\
J062224.23+032520.2 & 95.600993 & 3.422284 & 28 & G2 & Binary & new & NULL \\
J072058.63+250006.4 & 110.244321 & 25.001782 & 95 & Carbon & NULL & new & C-J(N) \\
\hline\hline
\end{tabular}
\end{center}
Note. The machine-readable version includes ultraviolet, optical, and infrared magnitudes and proper motion information. The format is similar to that shown in Tables 6--8.

(This table is available in its entirety in a machine-readable form.)
\end{table*}
\end{landscape}

\clearpage

\begin{table*}
\begin{center}
\caption{Numbers of C-H, C-R, C-J, C-N, and Barium Stars, Which have the $GALEX$, Pan-STARRS, 2MASS and $WISE$ Photometric Magnitudes.}\label{table:spectra_type_GALEX}
\begin{tabular}{cccccccc}
\hline\hline
Survey & Total Number \tablenotemark{a} & C-H & C-R & Ba & C-J & C-N & Unknown\\
\hline
$GALEX$ & 1099 & 370 & 109 & 262 & 41 & 12 & 47 \\
Pan-STARRS & 2608 & 852 & 223 & 712 & 387 & 259 & 176  \\
2MASS & 2567 & 852 & 225 & 715 & 359 & 247 & 158 \\
$WISE$ & 2550 & 850 & 222 & 713 & 352 & 241 & 161 \\
\hline\hline
\end{tabular}
\end{center}
\tablenotetext{a}{The total number of carbon star that have $GALEX$, Pan-STARRS, 2MASS, or $WISE$ photometric information.}
\end{table*}

\clearpage

\begin{table*}
\begin{center}
\caption{$GALEX$ Detections of 25 Binary Candidates that Likely Have Compact White Dwarf Companions.}\label{table:GALEX}
\begin{tabular}{c|c|c|c|c|c}
\hline\hline
Designation & fuv & fuv$\_$err  & nuv   & nuv$\_$err  & SpType \\
& (mag) & (mag) & (mag) & (mag) \\
\hline
J005749.75+013835.2 & 22.14 & 0.10 & 21.99 & 0.16 & C-H \\
J020726.72+453216.9 & 20.92 & 0.34 & 20.51 & 0.28 & C-N \\
J050736.14+305149.6 & 21.71 & 0.53 & -999 & -999 & Ba \\
J073406.93+351345.5 & 22.56 & 0.33 & 17.54 & 0.02 & Ba \\
J074743.32+173302.0 & 22.41 & 0.44 & 20.64 & 0.14 & C-H \\
J080917.58+004256.5 & 19.74 & 0.14 & 19.33 & 0.08 & Ba \\
J083021.22+154319.6 & 23.08 & 0.28 & 22.21 & 0.17 & UNKNOWN \\
J084906.99+462727.2 & 22.21 & 0.16 & 21.81 & 0.08 & UNKNOWN \\
J091555.05+043115.6 & 21.53 & 0.32 & 20.42 & 0.18 & Ba \\
J093450.24+022355.0 & 22.35 & 0.40 & 21.47 & 0.22 & Ba \\
J094634.19+140521.7 & 24.01 & 0.30 & 20.10 & 0.02 & C-H \\
J101110.08+285036.0 & 22.12 & 0.37 & 21.24 & 0.27 & Ba \\
J101423.40+302200.3 & 25.20 & 0.25 & 21.07 & 0.01 & C-H \\
J101946.87+252932.7 & 22.45 & 0.41 & 20.32 & 0.11 & C-R \\
J115932.16+014326.9 & 22.54 & 0.50 & 20.90 & 0.18 & C-H \\
J130359.18+050938.6 & 23.74 & 0.27 & 21.98 & 0.07 & C-H \\
J130824.28+530224.4 & 23.03 & 0.20 & 22.42 & 0.13 & C-H \\
J131525.84+062520.9 & 19.09 & 0.12 & 18.61 & 0.06 & C-H \\
J133841.23+014523.7 & 22.89 & 0.26 & 21.79 & 0.15 & UNKNOWN \\
J140953.08-061141.8 & 21.36 & 0.33 & 19.79 & 0.12 & C-H \\
J142057.12-031953.2 & 19.09 & 0.13 & 18.66 & 0.06 & C-H \\
J154903.86+033253.1 & 22.38 & 0.33 & 22.10 & 0.18 & UNKNOWN \\
J164420.62+034506.6 & 19.75 & 0.12 & 19.59 & 0.08 & C-J(UNKNOWN)\\
J212426.82-030344.6 & 23.98 & 0.42 & 21.53 & 0.10 & C-R\\
J220255.21-010708.3 & 21.12 & 0.06 & 21.00 & 0.05 & C-H\\
\hline\hline
\end{tabular}
\end{center}
\end{table*}

\clearpage

\newpage
\begin{appendix}

\section{The selection effect of positive samples}
In Section 2.1, we use four steps to select positive samples. At first, of the 1682 SDSS and LAMOST carbon stars, a fraction were observed multiple times. For such stars, we only retain the spectrum with the highest S/N, and exclude other spectra. Through this first step, we removed 234 repeatedly observed spectra. Next, we limit the wavelength range in step 2, and 13 spectra that did not satisfy our wavelength restriction are excluded by this step. In step 3, we remove 101 spectra which have abnormal fluxes, such as 0 or negative fluxes, or other problems. After these steps, there were still 1344 spectra left. When we manually checked those spectra, we find that 284 of them have considerable noise, making it difficult for us to identify them as carbon stars, and thus we only retain 1050 spectra, which we can definitely identify as carbons stars from spectral features, to construct a positive sample set. We remove the 284 ambiguous spectra from the positive sample set.

Figure~\ref{snr_all} shows the $r$-band S/N distribution of 1334 carbon star spectra, which are left after the third step as mentioned above. Figure~\ref{snr_ours} displays the S/N distribution of the 1050 positive samples used in this paper, and Figure~\ref{snr_low} exhibits the distribution of the 284 removed samples. From the three figures, we can clearly see that spectra with S/Ns lower than 10 are also included in our 1050 positive samples, which roughly accounts for less than 30 $\%$ of the entire samples. Unlike the 284 samples removed due to low-S/N spectra, those low-S/N positive samples that were retained have definitely distinguishable carbon molecular band features in their spectra. In addition, we can also see that the S/Ns of almost all of the 284 removed spectra are between 2 and 9, which indicates that these spectra have extremely low S/Ns.

Then, it is necessary to design an experiment to analyze the effect of the exclusion of these 284 low-S/N spectra on the recall. Firstly, we mix the 2651 carbon stars, which were found in Section 3 of this paper, into 500, 000 LAMOST spectra, and make them the unlabeled sample set U. Then, we implement two carbon star retrieval experiments. In the first experiment, we use the 1334 spectra mentioned above to construct the positive sample set P and retrieve carbon stars from set U. In the second experiment, we use the spectra of the 1050 positive samples used in this paper to construct set P, and also retrieve carbon stars from the set U. Finally, we compare the recall at different value of $K$ in the two experiments. Figure~\ref{snr_recall} presents the experiment results. In this figure, the magenta and black lines respectively present the results of the first and second experiments.

From Figure~\ref{snr_recall} , we can conclude that the recall of the second experiment (R2) is much higher than that of the first experiment (R1) when $K$ is smaller than 2000. However, when $K$ is larger than 2000, R1 is slightly higher than R2. For example, when $K = 3000$,  R1 is about 0.92$\%$ higher than R2, and R1 is about 0.99$\%$ higher than R2 at $K = 5000$. Therefore, if we do not classify the 1050 positive samples and use all of them to search for carbon stars, such an experiment result tells us that we may lose at least about 0.99$\%$ carbon stars at $K = 5000$. But, we should recognize that the small number of positive samples will make carbon star retrieval faster, although we may miss few carbon stars. In section 3, we search for carbon stars three times, and use one group of positive samples each time. Comparing Figure~\ref{snr_recall} with Figure~\ref{preprocess_I}--\ref{preprocess_III} in sub-section 3.1, we can clearly see that the recall is much higher if we classify the positive samples into several groups, which indicates our classification of positive samples in sub-section 2.2 is an extremely essential work. In addition, another investigation should be undertaken in subsequent work, and such a work will reveal whether we can get a higher recall than this paper if we classify 1344 positive samples into different groups and  search for carbon stars within each group.

\setcounter{figure}{0}
\renewcommand{\thefigure}{A.\arabic{figure}}
\begin{figure}
\centering
\epsscale{1.0}
\plotone{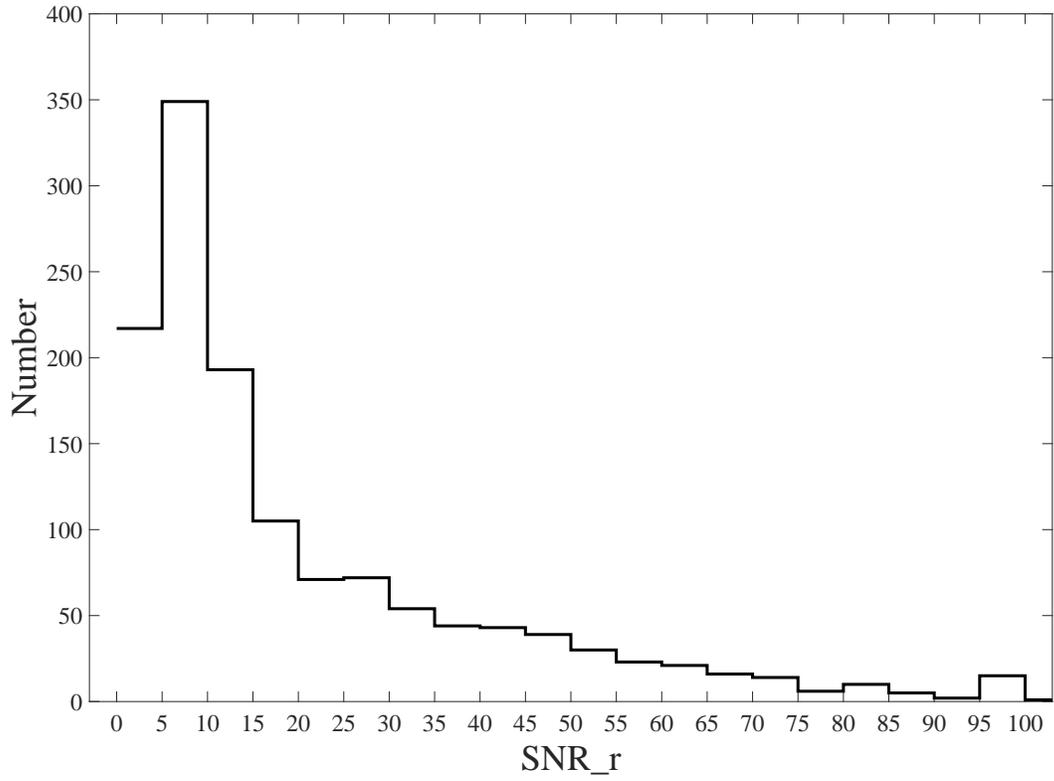}
\caption{S/N distribution of the SDSS $r$-band for 1334 carbon star spectra, which are retained after the step 3 of in Section 2.1 and include 284 low-S/N spectra. \label{snr_all}}
\end{figure}

\clearpage

\setcounter{figure}{1}
\renewcommand{\thefigure}{A.\arabic{figure}}
\begin{figure}
\centering
\epsscale{1.0}
\plotone{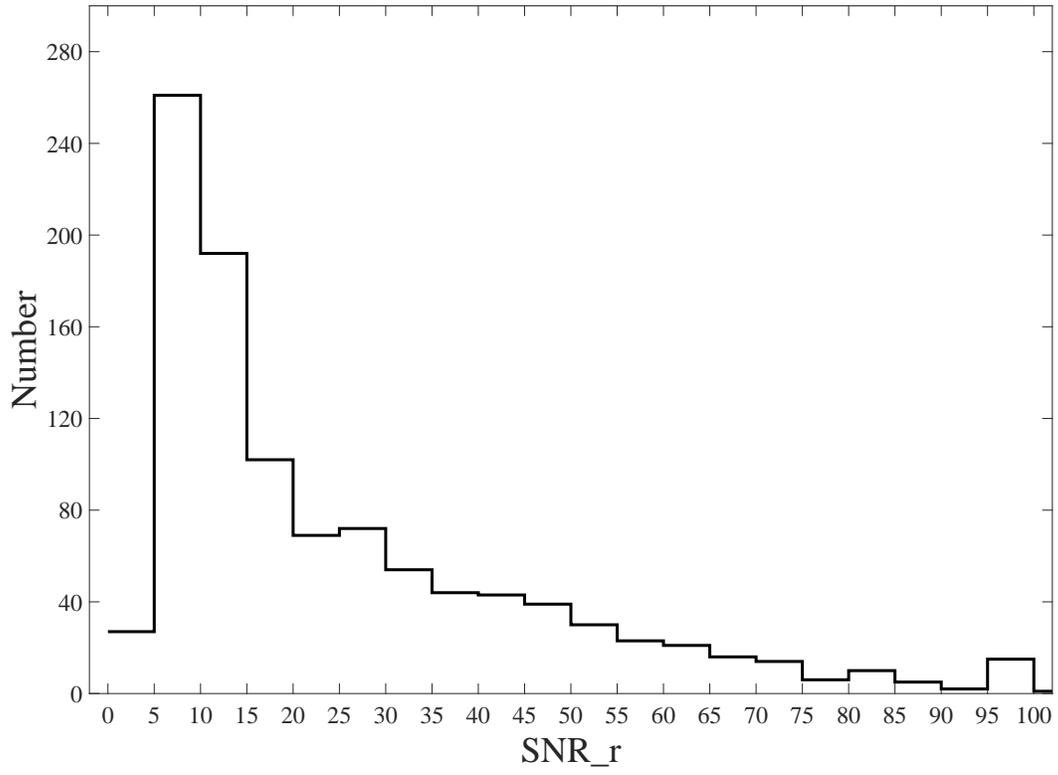}
\caption{S/N distribution of the SDSS $r$-band for the spectra of the 1050 positive samples, which are used to construct the positive sample set P in this paper. \label{snr_ours}}
\end{figure}

\clearpage

\setcounter{figure}{2}
\renewcommand{\thefigure}{A.\arabic{figure}}
\begin{figure}
\centering
\epsscale{1.0}
\plotone{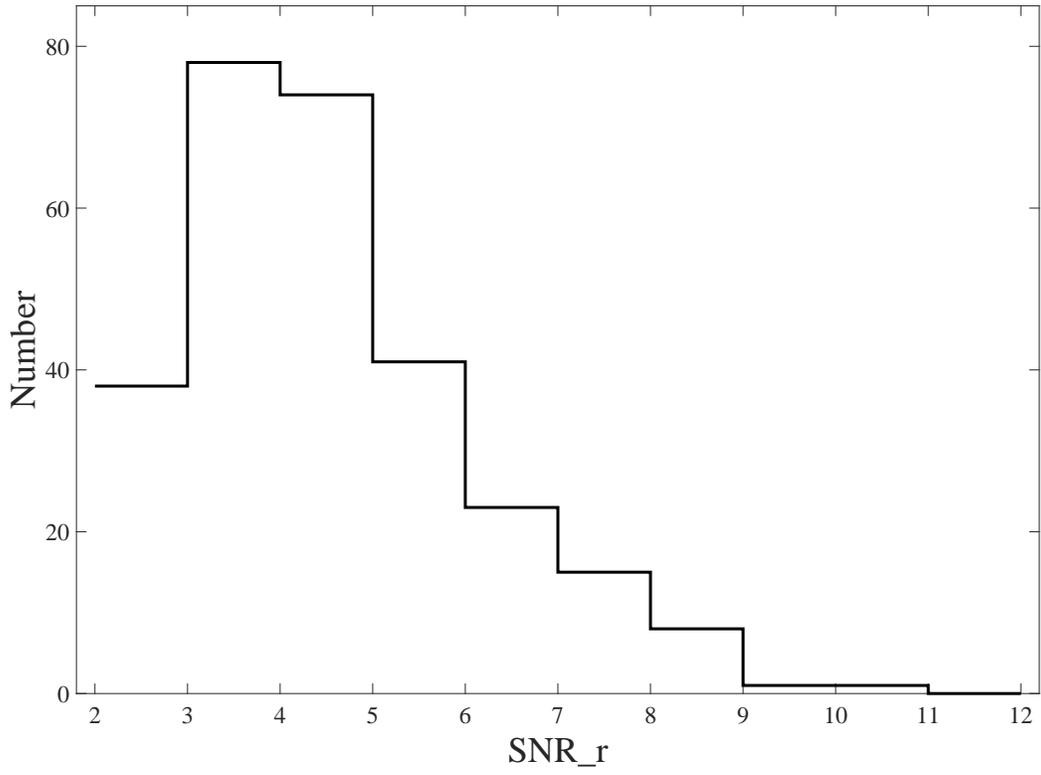}
\caption{S/N distribution of the SDSS $r$-band for the 284 low-S/N spectra, which are removed from the positive samples through step 4 in Section 2.1. \label{snr_low}}
\end{figure}

\clearpage

\setcounter{figure}{3}
\renewcommand{\thefigure}{A.\arabic{figure}}
\begin{figure}
\centering
\epsscale{0.9}
\plotone{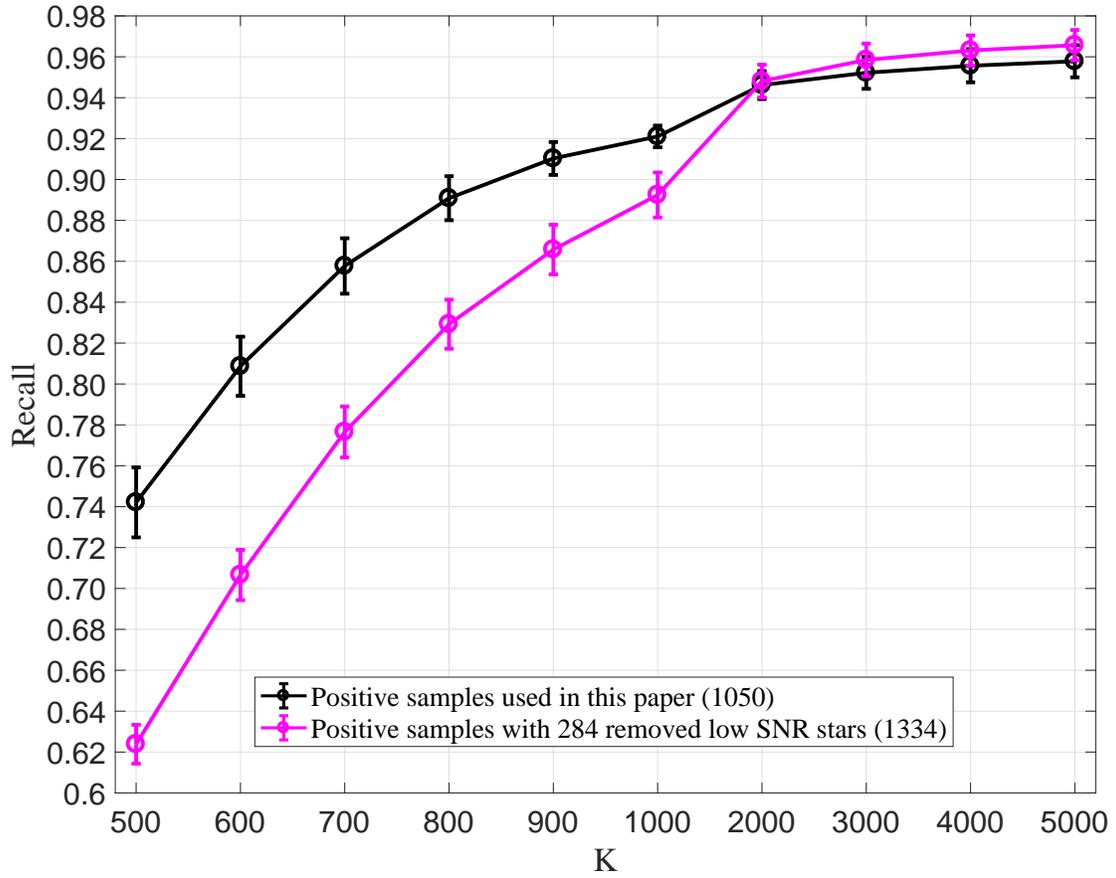}
\caption{Recall comparison.The black line shows the recall at different values of $K$ when we use the 1050 positive samples to search for carbon stars, and the magenta
line displays the recall when we use the 1334 positive samples, which include the 284 removed spectra, to search carbon stars. \label{snr_recall}}
\end{figure}

\clearpage

\section{Spectral Preprocessing Experimennts}

Generally, before searching for carbon stars from a massive data set, spectral data should be preprocessed firstly to get a better retrieval performance. The widely used preprocessing methods include de-noising, normalization, feature selection, and so on. In this paper, we firstly eliminate the disturbance of narrow lines, such as sky emission lines and bad pixels. Figure ~\ref{spectra_denoise} displays an example of spectra de-noising, and we can see that the bad pixels around 5800 \AA~have been removed effectively after filtering, showing that spectral de-nosing is an indispensable step. Second, it is necessary to normalize the spectral flux to the same scale. We try two commonly used normalization methods to check which method is more efficient. The first method maps the minimum and maximum fluxes of each spectra to [0, 1], which is abbreviated as ``nmap'' in this paper, and the other method normalizes the flux to the value of the square root of the flux square sum, which is marked as ``nunit.'' Third, we perform a median filter method with a width of 300 \AA\ to determine the pseudo-continuum, which is used to investigate the effect of the continuum on the retrieval performance. Finally, when searching for rare objects, we are often confronted with high dimensional spectral data, which may contain non-informative or noisy features, so it is necessary to extract the main information hidden in the spectral data. We apply the PCA, which has been popularly used to obtain the low-dimensional data representation, and 50 principal components have been retained.

Based on above preprocessing steps, we can obtain eight spectral preprocessing methods, which are referred to as ``nmap,'' ``nunit,'' ``nmap + pca,'' ``nunit + pca,'' ``nmap + subcon,'' ``nunit + subcon,'' ``nmap + pca + subcon,'' and ``nunit + pca + subcon.'' ``nmap'' and ``nunit'' indicate that spectra are preprocessed only by the ``nmap'' and  ``nunit'' normalizations; ``nmap + pca'' and ``nunit + pca'' indicate that spectra are preprocessed with the normalization and PCA dimension reduction, the ``nmap + subcontinuum'' and ``nunit + subcontinuum'' are two preprocesses with normalization and continuum subtraction, and the ``nmap + pca + subcontinuum'' and ``nunit + pca + subcontinuum'' indicate spectra are preprocessed by normalization, PCA dimension reduction, and continuum subtraction.

\setcounter{figure}{0}
\renewcommand{\thefigure}{B.\arabic{figure}}
\begin{figure}
\centering
\epsscale{1.0}
\plotone{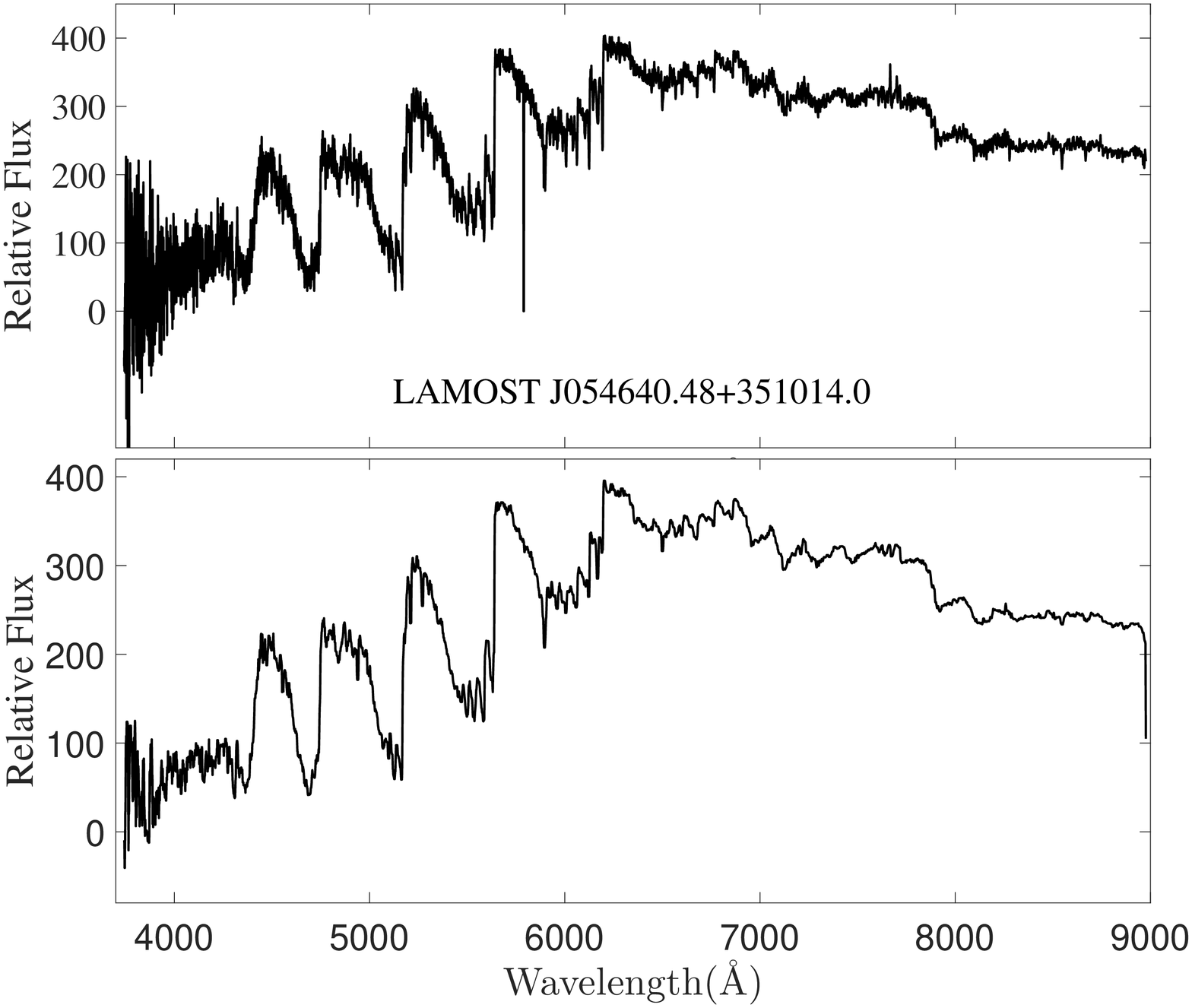}
\caption{The upper panel is the original spectrum of the object J054640.48+351014.0, and the bottom plot is the spectrum after filtering strong noises. \label{spectra_denoise}}
\end{figure}

\end{appendix}

\end{document}